\documentclass{article}

\usepackage[english]{babel}

\usepackage[letterpaper,top=2cm,bottom=2cm,left=3cm,right=3cm,marginparwidth=1.75cm]{geometry}

\usepackage{amsmath}
\usepackage{bm}
\usepackage{braket}
\usepackage{graphicx}
\usepackage{subcaption}
\usepackage[colorlinks=true, allcolors=blue]{hyperref}
\usepackage{authblk}
\usepackage{booktabs}
\usepackage{array}
\usepackage{tabularx}
\usepackage{rotating}
\usepackage{ragged2e} 
\usepackage[super]{nth}
\usepackage{float}
\usepackage{multirow}
\usepackage{tikz}
\usepackage{xcolor}

\newcolumntype{Y}{>{\RaggedRight\arraybackslash}X}
\newcolumntype{C}[1]{>{\centering\arraybackslash}p{#1}}

\def\checkmark{\tikz\fill[scale=0.4](0,.35) -- (.25,0) -- (1,.7) -- (.25,.15) -- cycle;}
\def\xmark{%
  \tikz[scale=0.4, baseline=(current bounding box.center)]{%
    \fill (0.05,0) -- (0.15,0)
          -- (0.7,0.55) -- (0.6,0.65) -- cycle;
    \fill (0.6,0) -- (0.7,0.1)
          -- (0.15,0.65) -- (0.05,0.55) -- cycle;
  }%
}

\begin{document}

\title{Benchmarking Quantum Neural Networks Developed for MBHB Coalescence in the Context of LISA Space Mission}

\author[1,2]{Maria-Catalina Isfan}
\author[1]{Laurentiu-Ioan Caramete}
\author[1]{Ana Caramete}

\affil[1]{Institute of Space Science – INFLPR Subsidiary, 409 Atomistilor, Magurele, Romania}
\affil[2]{Doctoral School of Physics, Faculty of Physics, University of Bucharest, 405 Atomistilor, Magurele, Romania}

\maketitle

\begin{abstract}
Machine learning (ML) techniques are increasingly being explored as tools for the analysis of gravitational wave (GW) data expected from the Laser Interferometer Space Antenna (LISA). At the same time, advances in quantum computing have stimulated interest in quantum machine learning as a potential alternative to conventional approaches. In this work, we investigate the applicability of quantum, hybrid quantum--classical, and classical ML models to two tasks relevant for LISA observations of massive black hole binaries (MBHB): GW signal detection and merger prediction. Using the Sangria dataset from the LISA Data Challenge, we benchmark several quantum architectures, including variational quantum classifiers (VQCs), post VQC hybrid models, quantum kernel methods, data reuploading classifiers, and quantum long short-term memory networks (QLSTMs), against classical support vector machines (SVMs), multilayer perceptrons (MLPs), recurrent neural networks (RNNs), and long short-term memory networks (QLSTMs). The models are trained on labelled data derived from the Sangria training dataset and evaluated on the corresponding blind dataset. For signal detection, several quantum models achieve performance comparable to or exceeding that of the classical baselines. In particular, the best performing hybrid post VQC architecture correctly identifies all signal samples in the blind dataset, while the standard VQC and the data reuploading classifier achieve competitive detection rates. For merger detection and coalescence time estimation, however, the standard VQC emerges as the most reliable model, successfully localising all detected merger events and producing the smallest localisation errors. We further analyse the trade-offs between predictive performance, model complexity, and quantum-circuit resources. Our results indicate that, although no universal quantum advantage is observed, selected quantum models can provide practical benefits for specific gravitational-wave analysis tasks and represent promising candidates for future LISA data analysis pipelines.
\end{abstract}

\section{Introduction}

The Laser Interferometer Space Antenna (LISA) \cite{LISA} is expected to open a new observational window on the milli-Hertz gravitational wave (GW) Universe. Among its most promising sources are massive black hole binaries (MBHBs), whose inspiral, merger, and ringdown phases provide unique information about black hole growth, galaxy evolution, and tests of gravitation in the strong field regime. Owing to the long inspiral phase observable by LISA, MBHB signals may be detected weeks to months before merger, creating opportunities for multimessenger observations and low latency alerts \cite{multimesseger}. Achieving these goals, however, requires reliable detection and rapid characterisation of signals embedded in complex instrumental and astrophysical noise. Recent studies have therefore focused on the development of low-latency methods for the early detection and preliminary characterisation of MBHBs in LISA data streams, including time-frequency approaches and machine-learning-based methodologies \cite{Nobili2026,Spadaro2026}. In particular, future low-latency analysis pipelines will need to identify potential GW signals shortly, within 1 hour, after the data become available for scientific analysis, enabling prompt alerts and follow-up observations prior to coalescence \cite{LISA}.

To address these challenges, machine learning (ML) techniques have become increasingly important in GW data analysis \cite{GW_ML}. Classical support vector machines (SVMs) \cite{GW_SVM} and deep learning architectures \cite{GW_deepfiltering} have been applied to signal detection, denoising, source classification, and parameter estimation. Their ability to learn directly from data offers an attractive alternative to computationally expensive matched filtering pipelines, particularly when analysing large datasets expected from next generation observatories.

In parallel, rapid developments in quantum computing have stimulated interest in quantum machine learning (QML) \cite{QML}. Variational quantum algorithms \cite{VQAs}, quantum kernel methods, quantum-enhanced recurrent networks, and hybrid quantum-classical architectures have been proposed as potential alternatives to conventional learning models. Although current quantum hardware remains limited \cite{NISQ}, numerous papers suggest that quantum representations may provide expressive feature spaces capable of capturing complex correlations in scientific datasets \cite{q_features_1} \cite{q_features_2} \cite{q_data}. However, whether such models offer practical benefits for realistic astrophysical applications remains an open question. This motivates the exploration of GW data analysis as a use case for QML, where realistic datasets, well defined detection tasks, and established classical baselines provide a framework for assessing the practical value of quantum learning approaches.

In our previous work \cite{VQC_ul_nostru}, we investigated the application of VQCs to the detection of MBHB signals in data from the Sangria LISA Data Challenge. Using the Sangria dataset \cite{sangria}, we demonstrated that VQC-based QNN architectures achieved competitive detection performance and successfully identified GW signals in previously unseen data. Those results established the feasibility of applying QML techniques to realistic LISA datasets and motivated a more systematic investigation of their capabilities and limitations \cite{articol_model}. However, signal detection alone is insufficient for many future LISA science applications. For low-latency analyses and multimessenger follow-up campaigns \cite{multimesseger}, it is equally important to identify merger candidates and estimate their coalescence times ($t_c$). Furthermore, the relative performance of different quantum architectures, as well as their comparison with established classical approaches, remains largely unexplored. In particular, it is not yet clear whether the improved flexibility of quantum and hybrid models translates into better generalisation, more accurate merger localisation, or improved computational efficiency.

The primary goal of the present work is therefore twofold. First, we perform a comprehensive benchmarking study of quantum, hybrid quantum-classical, and classical NN models for MBHB signal detection in the Sangria LISA Data Challenge dataset \cite{sangria}. Second, we extend the analysis beyond signal detection and investigate merger detection and $t_c$ estimation. The considered quantum models include VQCs \cite{VQC}, post VQC hybrid architectures \cite{post_VQC}, data reuploading classifiers \cite{data_reuploading_classifier}, quantum kernel methods \cite{q_kernel_1} \cite{q_kernel_2}, and quantum long short-term memory (QLSTM) networks \cite{QLSTM_paper_1} \cite{QLSTM_paper_2}, which are benchmarked against classical SVMs \cite{SVM_original_paper}, multilayer perceptrons (MLPs) \cite{MLP_original_paper}, RNNs \cite{RNN_original_paper}, and long short-term memory (LSTM) networks \cite{LSTM_original_paper}.

\section{The Dataset}

\begin{figure}
    \centering
    \includegraphics[width=\linewidth]{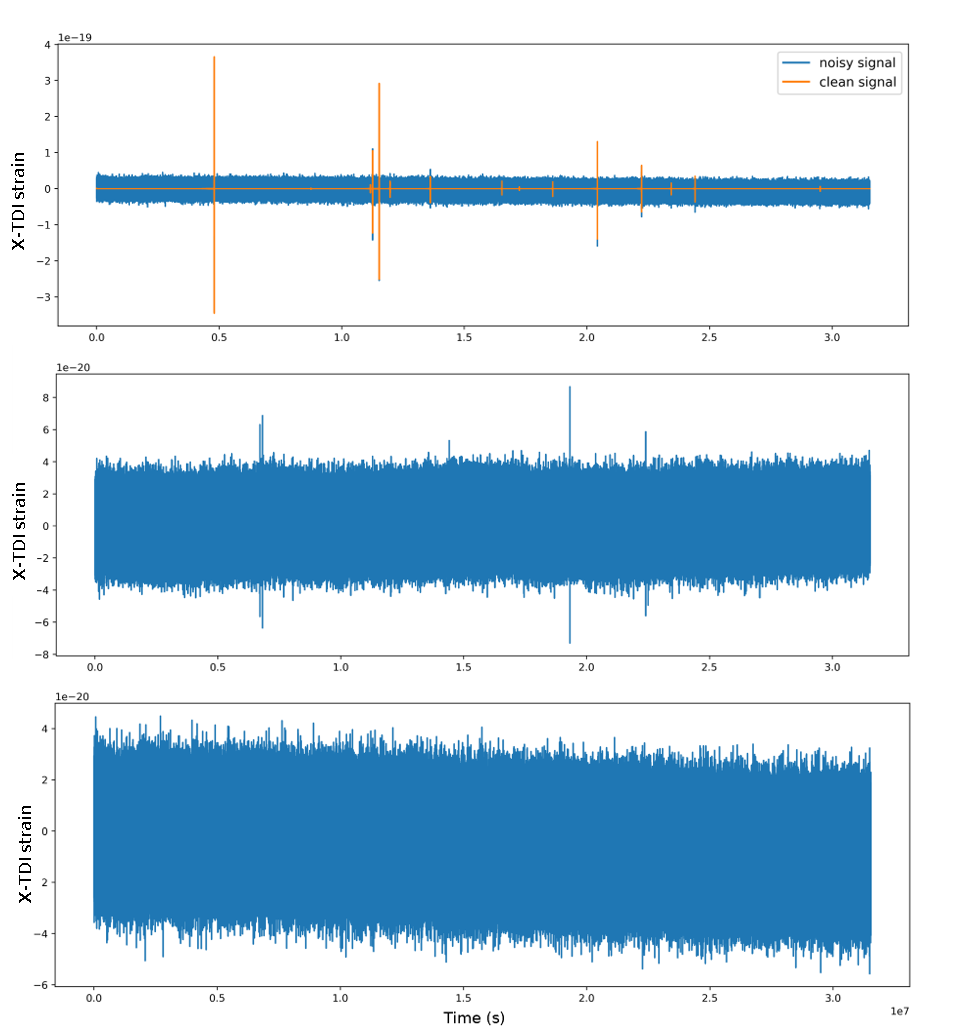}
    \caption{Upper panel: the training dataset consisting of 15 MBHB GW consecutive signals. Middle panel: the blind dataset, with undisclosed information. Lower panel: noise in the training dataset.}
    \label{fig:data_sets}
\end{figure}

We tailor and evaluate our NN models on the Sangria LISA Data Challenge \cite{sangria}. The challenge involves studying a dataset with known full specification used to simulate it and subsequently analyzing a blind dataset with undisclosed information. Each are one year time series GW strain generated within the LDC group of the LISA Consortium \cite{LDC} using the LDC software \cite{LDCsoftware}. In the following, we refer to the former as the \textit{training dataset} and to the later as the \textit{blind dataset}. In our approach, we train the models on the training dataset and perform inference on the blind dataset. The train dataset (Figure \ref{fig:data_sets}, upper panel) consists of 15 consecutive MBHB GW signals embedded in intricate noise, including verification Galactic binaries waveforms, Galactic white dwarf binaries waveforms and  confusion foreground (Figure \ref{fig:data_sets}, lower panel). For this dataset, the clean (noiseless) signal is also provided (Figure \ref{fig:data_sets}, upper panel), allowing to isolate the noise by substracting the clean signal from the train dataset. 

\subsection{Our Approach to GW Data Analysis Using QML}

Our goal is to rapidly detect MBHB GW signals in LISA data streams and provide a minimal characterisation of the signal, such as estimates of the coalescence time $t_c$. As in our previous work \cite{VQC_ul_nostru}, we develop QNNs and quantum–classical hybrid NNs to (i) classify GW data samples as either signal or noise, and (ii) perform merger event prediction and coalescence time estimation for the Sangria LISA Data Challenge \cite{sangria}. The ultimate goal of this work is to benchmark the QNNs and the quantum-classical hybrid NNs against classical NNs and evaluate the quantum advantage of the fore methods.

For (i), we train the models on the training dataset, using both the signal+noise data and the noise obtained by subtracting the provided clean signal, and perform inference on the blind dataset, aiming to classify its samples as noisy GW signals. For (ii), we train on the training dataset and subsequently identify merger events in the blind dataset, while also estimating $t_c$.

These two analysis modules can be used independently or jointly for consistency checks. Each module requires a distinct data pre-processing strategy.

\subsection{Data Pre-processing}

To feed one-year time-series data into the NN models (quantum and classical), we apply a three-step pre-processing procedure for each module, in the same manner as in our earlier work \cite{VQC_ul_nostru}: (i) slicing the data to generate multiple samples, (ii) assigning labels to each sample, and (iii) mapping each sample to its corresponding feature representation. Selected features are then used as input to the NN models.

\subsubsection{Data Slicing and Samples Generation}

We are provided with one-year time-series. Data in this form is not suitable for processing with NNs, as these algorithms require multiple data samples to train on. To overcome this, we implement a moving window approach \cite{movingwindows1} \cite{movingwindows2} to slice the data and generate multiple data samples. So, for the first module of our analysis, we slice the training dataset, the isolated noise dataset according and the blind dataset to Eq.(\ref{eq:noofadjacentwindows}) and generate 1802 adjacent data samples per dataset. For the first module, we slice the training dataset and the blind dataset according to Eq.(\ref{eq:noofmovingwindows}), generating 63062 data samples,with samples overlapping for 500 seconds.

\begin{equation}
    \text{no. of samples} = \left\lceil \frac{\text{no. of data points}}{\text{window length}} \right\rceil
    \label{eq:noofadjacentwindows}
\end{equation}

\begin{equation}
    \text{no. of samples} = \frac{\text{no. of data points} - \text{window length}}{\text{step}}
    \label{eq:noofmovingwindows}
\end{equation}

\subsubsection{Labeling}

For signal detection, we label the samples derived from the training dataset as ‘1’, corresponding to noisy GW signals, and the samples derived from the isolated noise dataset as ‘0’, corresponding to noise without GW signals. For the blind dataset, we treat the samples as potential noisy GW signals and evaluate the models’ ability to classify them accordingly, i.e. to perform GW signal detection on unseen data.

For merger event prediction, we adopt a more elaborate labeling strategy. We first assign a label to each point in the training dataset, obtaining a label array with the same length as the training time series. We then slice this label array according to Eq.(\ref{eq:noofmovingwindows}), yielding the same number of label windows as data samples, with each label window associated with a corresponding data sample.
Since a label window may contain both 0s and 1s, we map it to a single label by assigning it a value of 1 if it contains at least one 1, and 0 otherwise.r it as 0.

\begin{figure}
    \centering
    \includegraphics[width=1\linewidth]{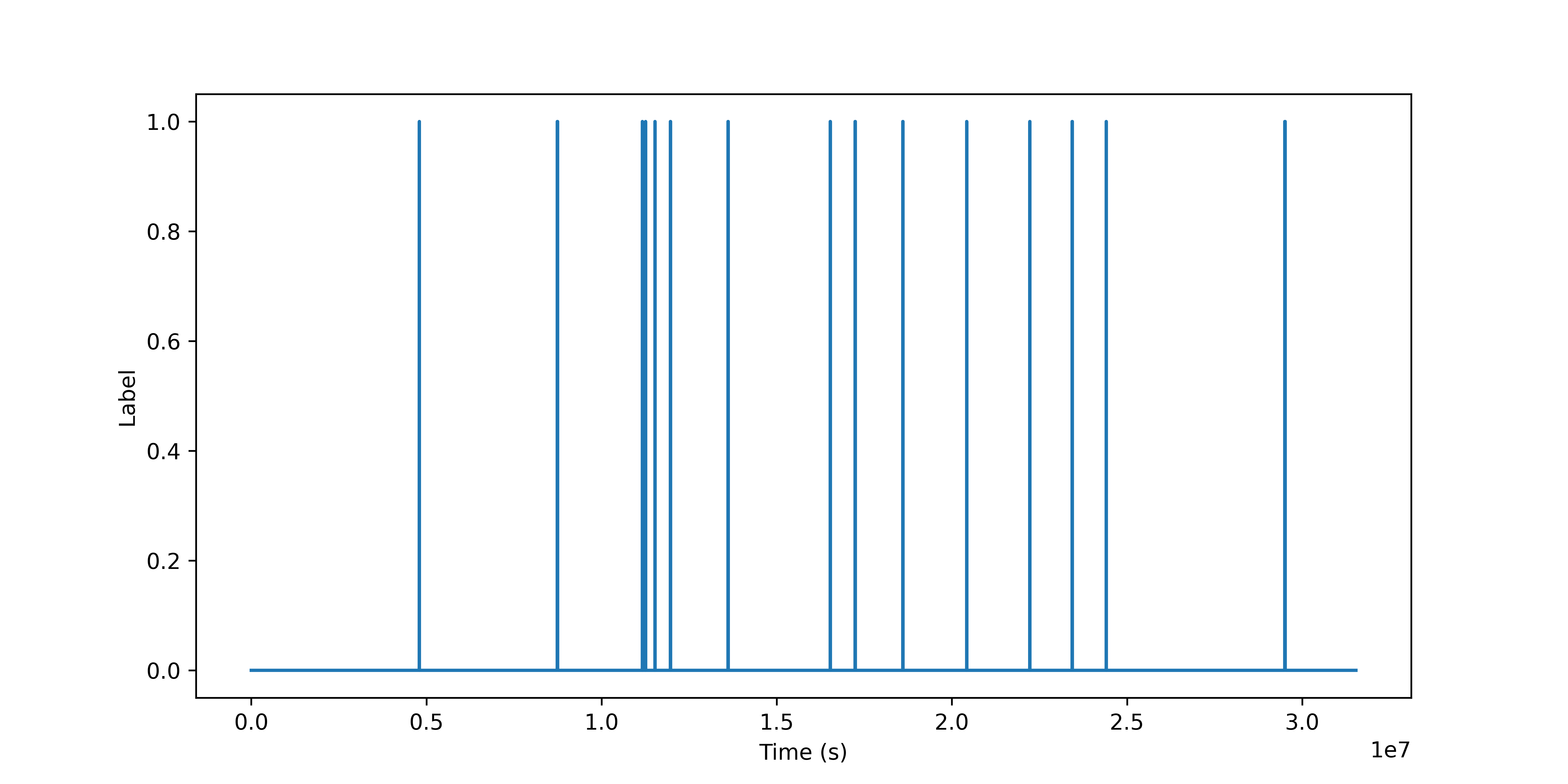}
    \caption{Caption}
    \label{fig:labels_ISCO_each_point_VS_time}
\end{figure}

The methodology for point-by-point label assignation is as follows. We adopt a simplified approach, starting from the frequency of the emitted GWs by inspiraling MBHBs in the Newtonian approximation (Eq.(\ref{eq:freq})) \cite{Maggiore} and working in the frozen low-frequency approximation \cite{freq_approx}. When the MBHB reaches its innermost stable circular orbit (ISCO), the emitted GWs have the frequency given in Eq.(\ref{eq:f_ISCO}). The corresponding time at which this occurs is $t_{ISCO}$ (Eq.(\ref{eq:f_ISCO_t_ISCO})). By inverting Eq.(\ref{eq:f_ISCO_t_ISCO}), we obtain an expression for $t_{ISCO}$ (Eq.(\ref{eq:t_ISCO})).
Since the training dataset provides the masses and $t_c$ for each GW source, we can compute $f_{ISCO}$ and $t_{ISCO}$ for each source. We then assign a label of 1 to all time instances between $t_{ISCO}$ and $t_c$, and 0 otherwise (Figure \ref{fig:labels_ISCO_each_point_VS_time}). 

\begin{equation}
    f( t, t_c, \mathcal{M}_c ) = \frac{1}{\pi} \Big( \frac{2}{256} \frac{1}{t_c - t} \Big) ^\frac{3}{8} \Big( \frac{G \mathcal{M}_c}{c^3} \Big) ^{-\frac{5}{8}}
    \label{eq:freq}
\end{equation}

\begin{equation}
    f_{ISCO} = \frac{1}{6\sqrt{6}(2\pi)} \frac{c^3}{GM}
    \label{eq:f_ISCO}
\end{equation}

\begin{equation}
    f_{ISCO}( t_{ISCO}, t_c, \mathcal{M}_c ) = \frac{1}{\pi} \Big( \frac{2}{256} \frac{1}{t_c - t_{ISCO}} \Big) ^\frac{3}{8} \Big( \frac{G \mathcal{M}_c}{c^3} \Big) ^{-\frac{5}{8}}
    \label{eq:f_ISCO_t_ISCO}
\end{equation}

\begin{equation}
    t_{ISCO} = t_c - \frac{2}{256} ( \pi f_{ISCO} ) ^{-\frac{8}{3}} \Big( \frac{c^3}{G\mathcal{M}_c} \Big) ^{-\frac{5}{3}} 
    \label{eq:t_ISCO}
\end{equation}

\subsubsection{Features computation}

We map each data sample to a corresponding signal processing feature vector \cite{signealprocessingfeatures}, regardless the data analysis module and preserving the labels. In this way, we represent time-series data as compact, selected feature vectors, which are then fed into our models.

\section{Quantum Neural Network Models Implemented for our Analysis and Benchmarking} \label{sec:quantummodels}

We consider a set of established quantum, classical, and quantum–classical hybrid NN models. The quantum models include VQC \cite{VQC, VQC_1, VQC_2, VQC_pennylane_demo} and the data re-uploading classifier \cite{data_reuploading_classifier, data_reuploading_classifier_pennylane_demo}. The classical models include SVM \cite{SVM_original_paper}, MLP \cite{MLP_original_paper}, RNN \cite{RNN_original_paper}, and LSTM \cite{LSTM_original_paper}. The hybrid models include post VQCs \cite{post_VQC, post_VQC_pennylane_demo}, quantum kernel SVMs \cite{q_kernel_1, q_kernel_2, q_kernel_1_pennylane_demo, q_kernel_2_pennylane_demo}, and QLSTM \cite{QLSTM_paper_1, QLSTM_paper_2, QLSTM_tutorial, QLSTM_tutorial_paper}.
The classical models follow standard Keras implementations \cite{Keras} and we therefore do not describe them in detail. We instead focus on the quantum and hybrid models.

\subsection{VQC}

The Variational Quantum Classifier (VQC) is a parameterized quantum model that combines a quantum feature map $U(\bm{x})$ with a trainable variational circuit $W(\bm{\theta})$ \cite{VQC, VQC_1, VQC_2, VQC_pennylane_demo}. The specific feature maps and ansätze employed in this work are summarized in Table~\ref{tab:quantum_models}.

Given an input feature vector $\bm{x}$, the quantum circuit produces the expectation value

\begin{equation}
    f(\bm{x}, \bm{\theta}) = \langle 0 | U^{\dagger}(\bm{x}) W^{\dagger}(\bm{\theta_1}) \cdots W^{\dagger}(\bm{\theta_n}) O W(\bm{\theta_n}) \cdots W(\bm{\theta_1}) U(\bm{x}) |0\rangle,
\end{equation}

where $O = Z \otimes I \otimes \cdots$ and only the first qubit is measured. The quantity $f(\bm{x},\bm{\theta})$ represents the raw output of the quantum circuit and satisfies $f(\bm{x},\bm{\theta}) \in [-1,1]$.

To map the circuit output to the interval $[0,1]$, we define

\begin{equation}
    p(\bm{x}, \bm{\theta}) = \frac{1 + f(\bm{x}, \bm{\theta})}{2}.
\end{equation}

The predicted class label is obtained by thresholding at $p = 0.5$,

\begin{equation}
    y_{\mathrm{pred}} = \frac{1}{2}\left(1 + \mathrm{sgn}\left(p(\bm{x}, \bm{\theta}) - 0.5\right)\right).
\end{equation}

The trainable parameters $\bm{\theta}$ are optimized by minimizing the square loss

\begin{equation}
    \ell(\bm{\theta}, \bm{x}, \bm{y}) = \left(f(\bm{x}, \bm{\theta}) - \bm{y}\right)^2.
\end{equation}

\subsection{Post VQC}

The post VQC model combines a VQC with a classical NN \cite{post_VQC, post_VQC_pennylane_demo}. The VQC acts as a quantum feature extractor by generating one or more quantum features from the input data, while the final classification is performed by the classical NN. The specific quantum feature extraction strategies employed in this work are summarized in Table~\ref{tab:quantum_models}.

Given an input feature vector $\bm{x}$, the VQC produces a feature vector $\bm{z}$ whose components are expectation values of one or more observables. These quantum features are subsequently processed by a classical NN

\begin{equation}
    g(\bm{z}) = \sigma \big( W_L \phi_{L-1}(\cdots \phi_1(W_1 \bm{z} + b_1)\cdots ) + b_L \big),
\end{equation}

where $\phi_i$, $i=\overline{1,L}$, denote activation functions, $\sigma(\bm{u}) = (1+e^{-\bm{u}})^{-1}$ is the sigmoid function, and $\bm{\omega}=\{W_i,b_i\}$, $i=\overline{1,L}$, are the trainable weights and biases of the NN.

The predicted class label is given by

\begin{equation}
    y_{\mathrm{pred}} = g(f(\bm{x}, \bm{\theta}), \bm{\omega}),
\end{equation}

where $f(\bm{x},\bm{\theta})$ denotes the quantum feature extraction stage. The trainable parameters $\bm{\theta}$ and $\bm{\omega}$ are jointly optimized using the binary cross-entropy loss

\begin{equation}
    \mathcal{L}(\bm{\theta}, \bm{\omega}) = \frac{1}{N} \sum_{i=1}^{N} \ell(x_i,y_i;\bm{\theta},\bm{\omega}),
\end{equation}

with

\begin{equation}
    \ell(x_i,y_i;\bm{\theta},\bm{\omega}) = -\Big[y_i \log g(f(\bm{\theta},x_i),\bm{\omega}) + (1-y_i)\log\big(1-g(f(\bm{\theta},x_i),\bm{\omega})\big)\Big].
\end{equation}

\subsubsection{Observables construction heuristic}

In the observables construction heuristic, additional quantum features are generated by measuring observables of increasing locality while keeping the feature map fixed. The quantum feature extraction stage is

\begin{equation}
    f(\bm{x}, \bm{\theta}) = \braket{0|U^{\dagger}(\bm{x}) O_l U(\bm{x})|0},
\end{equation}

where

\begin{equation}
    O_l=\{P\in\mathcal{P}_n \mid w(P)=l\}
    \label{eq:obs_cirq}
\end{equation}

denotes an $l$-local observable and $w(P)$ is the number of non-identity Pauli operators in $P$. The resulting expectation values constitute the input feature vector $\bm{z}$ of the NN.

\subsubsection{Ansatz expansion heuristic}

In the ansatz expansion heuristic, additional quantum features are generated by evaluating the circuit for different fixed parameter configurations. The quantum feature extraction stage is

\begin{equation}
    f(\bm{x}, \bm{\theta}_j) = \braket{0|U^{\dagger}(\bm{x})W_o^{\dagger}(\bm{\theta}_j)OW_o(\bm{\theta}_j)U(\bm{x})|0},
\end{equation}

where $O=Z\otimes I$. The expanded ansatz $W_o(\bm{\theta}_j)$ is constructed by systematically shifting subsets of circuit parameters while keeping all remaining parameters fixed. The expectation values obtained from the resulting circuits form the feature vector $\bm{z}$ supplied to the NN.

\subsubsection{Observables construction heuristic + ansatz expansion heuristic}

The third variant combines the observables construction and ansatz expansion heuristics. Quantum features are generated using both higher-locality observables and fixed-parameter ansatz expansions,

\begin{equation}
    f(\bm{x}, \bm{\theta}_j) = \braket{0|U^{\dagger}(\bm{x})W_o^{\dagger}(\bm{\theta}_j)O_lW_o(\bm{\theta}_j)U(\bm{x})|0}.
\end{equation}

Here, $O_l$ is the $l$-local observable defined in Eq.~(\ref{eq:obs_cirq}) and $W_o(\bm{\theta}_j)$ is the expanded ansatz. The resulting expectation values are concatenated to form the feature vector $\bm{z}$ used by the NN.

\subsection{Data Reuploading Classifier}

The data reuploading classifier is a VQC architecture in which the input data are encoded multiple times throughout the circuit \cite{data_reuploading_classifier, data_reuploading_classifier_pennylane_demo}. By repeatedly applying the feature map between trainable circuit layers, the model can construct more expressive decision boundaries while using a limited number of qubits. The specific feature maps and ansätze employed in this work are summarized in Table~\ref{tab:quantum_models}.

Given an input feature vector $\bm{x}$, the quantum circuit produces the expectation value

\begin{equation}
    f(\bm{x}, \bm{\theta}) = \braket{ 0 | U^{\dagger}(\bm{x}) W^{\dagger}(\bm{\theta_1}) \cdots U^{\dagger}(\bm{x}) W^{\dagger}(\bm{\theta_n}) O U^{\dagger}(\bm{x}) W(\bm{\theta_n}) \cdots U^{\dagger}(\bm{x}) W(\bm{\theta_1}) U(\bm{x}) | 0 }.
\end{equation}

Unlike the standard VQC, where the feature map is applied once, the data reuploading classifier alternates data encoding and trainable layers throughout the circuit. This repeated encoding allows the model to capture more complex relationships between the input features and the target labels.

The observable is defined as

\begin{equation}
    O = \ket{y}\bra{y},
\end{equation}

where $y \in \{0,1\}$ denotes a class label. The fidelity between the quantum state produced by the circuit and the target label state is used as the classification score.

The predicted label is obtained according to

\begin{equation}
    y_{\mathrm{pred}} = \mathrm{argmax}\Big(F\big(\ket{\psi(\bm{x},\bm{\theta})},\ket{y}\big)\Big),
\end{equation}

where $F$ denotes the fidelity and $\ket{\psi(\bm{x},\bm{\theta})}$ is the output quantum state.

The trainable parameters $\bm{\theta}$ are optimized by minimizing the cost function

\begin{equation}
    C = \sum_{i=1}^{N}\Big(1 - F\big(\ket{\psi(\bm{x_i},\bm{\theta})},\ket{y_i}\big)\Big).
    \label{eq:datareup_cost}
\end{equation}

Minimizing Eq.~(\ref{eq:datareup_cost}) maximizes the fidelity between the output quantum states and their corresponding target label states, thereby improving the classification performance.

\subsection{Quantum Kernels}

Quantum kernel methods combine a quantum feature map with a classical kernel-based classifier \cite{q_kernel_1, q_kernel_1_pennylane_demo, q_kernel_2, q_kernel_2_pennylane_demo}. Instead of training a parameterized quantum circuit, these models evaluate similarities between data samples in a quantum feature space. The specific feature maps employed in this work are summarized in Table~\ref{tab:quantum_models}.

Given two input feature vectors, $\bm{x}_i$ and $\bm{x}_j$, their similarity is quantified through the quantum kernel

\begin{equation}
    \kappa(\bm{x}_i,\bm{x}_j) = \left| \bra{0} U^{\dagger}(\bm{x}_i) U(\bm{x}_j) \ket{0} \right|^2.
\end{equation}

The kernel measures the overlap between the quantum states associated with the two inputs and therefore defines a similarity metric in the quantum feature space induced by the feature map $U(\bm{x})$.

The resulting kernel matrix is subsequently used by a classical support vector machine (SVM), whose prediction is given by

\begin{equation}
    y_{\mathrm{pred}}(\bm{x}) = \mathrm{sgn} \left( \sum_{i=1}^{N} \alpha_i \kappa(\bm{x}_i,\bm{x}) + b \right),
\end{equation}

where $\alpha_i$ are trainable coefficients associated with the support vectors and $b$ is the bias term.

The model parameters are optimized using the hinge loss

\begin{equation}
    L(y_j,y_{\mathrm{pred}}(\bm{x}_j)) = \max\left(0,1-y_jy_{\mathrm{pred}}(\bm{x}_j)\right).
\end{equation}

\subsection{QLSTM}

The Quantum Long Short-Term Memory (QLSTM) model combines the classical LSTM architecture with parameterized quantum circuits \cite{QLSTM_paper_1, QLSTM_paper_2, QLSTM_tutorial, QLSTM_tutorial_paper}. The overall architecture follows the standard LSTM structure, while the forget, input, update, and output gates are implemented using quantum circuits. The specific feature maps and ansätze employed in this work are summarized in Table~\ref{tab:quantum_models}.

Given an input vector $\bm{z}$, a quantum gate is implemented as

\begin{equation}
    \bm{f}_{gate}(\bm{\theta}_{gate}, \bm{z}) = \bra{0} W^\dagger(\bm{z}) U^\dagger(\bm{\theta}_{gate}) O U(\bm{\theta}_{gate}) W(\bm{z}) \ket{0},
\end{equation}

where all qubits are measured and

\begin{equation}
    O = Z \otimes Z \otimes \cdots.
\end{equation}

The output of the quantum circuit is subsequently processed by a classical linear layer

\begin{equation}
    g_{gate}(\bm{x}) = \bm{x}A^T + b,
\end{equation}

where $A$ is the weights matrix and $b$ is the bias vector.

The QLSTM retains the standard forget, input, update, and output gate structure of the classical LSTM. These gates are computed according to

\begin{equation}
    f_t = \sigma \Big( g_f \big( \bm{f}_f \big( \bm{\theta}_f, [\bm{h}_{t-1}, \bm{x}_t] \big) \big) \Big),
    \label{eq:forget_gate}
\end{equation}

\begin{equation}
    i_t = \sigma \Big( g_i \big( \bm{f}_i \big( \bm{\theta}_i, [\bm{h}_{t-1}, \bm{x}_t] \big) \big) \Big),
    \label{eq:input_gate}
\end{equation}

\begin{equation}
    u_t = \tanh \Big( g_u \big( \bm{f}_u \big( \bm{\theta}_u, [\bm{h}_{t-1}, \bm{x}_t] \big) \big) \Big),
    \label{eq:update_gate}
\end{equation}

\begin{equation}
    o_t = \sigma \Big( g_o \big( \bm{f}_o \big( \bm{\theta}_o, [\bm{h}_{t-1}, \bm{x}_t] \big) \big) \Big),
    \label{eq:output_gate}
\end{equation}

where $[\cdot,\cdot]$ denotes vector concatenation.

As in a classical LSTM, the cell state and hidden state evolve according to

\begin{equation}
    c_t = f_t \odot c_{t-1} + i_t \odot u_t,
\end{equation}

\begin{equation}
    \bm{h}_t = o_t \odot \tanh(c_t).
\end{equation}

Here, $\odot$ denotes element-wise multiplication. The hidden state produced at the final time step is used for classification,

\begin{equation}
    y_{\mathrm{pred}} = \mathrm{argmax}(\bm{h}_t^{\mathrm{final}}).
\end{equation}

The trainable parameters $\bm{\theta}$ are optimized using the binary cross-entropy loss

\begin{equation}
    \mathcal{L}(\bm{\theta}) = \frac{1}{N} \sum_{i=1}^{N} \ell(x_i,y_i;\bm{\theta}),
\end{equation}

where

\begin{equation}
    \ell(x_i,y_i;\bm{\theta}) = -\Big[y_i\log g(f(\bm{\theta},x_i)) + (1-y_i)\log\big(1-g(f(\bm{\theta},x_i))\big)\Big].
\end{equation}

\begin{sidewaystable}[htbp]
\centering
\scriptsize
\setlength{\tabcolsep}{4pt}
\renewcommand{\arraystretch}{1.1}

\begin{tabular}{p{2.2cm}
p{2.2cm}
p{2.1cm}
p{1.9cm}
p{1.9cm}
p{2.3cm}
p{2.2cm}
p{2.7cm}
p{2.1cm}
p{1.3cm}
p{1.3cm}}
\toprule

\multirow{2}{*}{\textbf{Model}}
& \multicolumn{2}{c}{\textbf{GW signal features}}
& \multicolumn{2}{c}{\textbf{Embedding}}
& \multirow{2}{*}{\textbf{Ansatz}}
& \multirow{2}{*}{\textbf{Measurement}}
& \multirow{2}{*}{\textbf{Hyperparameters}}
& \multirow{2}{*}{\textbf{Classical processing}}
& \multicolumn{2}{c}{\textbf{Loss}} \\

\cmidrule(lr){2-3}
\cmidrule(lr){4-5}
\cmidrule(lr){10-11}

& \textbf{GW detection}
& \textbf{Merger detection}
& \textbf{GW detection}
& \textbf{Merger detection}
& & & &
& \textbf{GW detection}
& \textbf{Merger detection}
\\

\midrule

VQC
&
-- power spectral density mean\newline
-- standard deviation\newline
-- mean\newline
-- quantile 0.25\newline
-- quantile 0.75\newline
-- spectral entropy
&
-- peak frequency\newline
-- power spectral density mean\newline
-- standard deviation\newline
-- maximum\newline
-- spectral entropy
&
angle embedding with $R_X$ rotations
&
angle embedding with $R_X$ rotations
&
two-layer circuit with arbitrary rotations and cyclic CNOT gates
&
first-qubit Pauli Z
&
-- learning rate\newline
-- batch size\newline
-- n layers\newline
-- n epochs
&
trainable bias added to output of circuit
&
square
&
square
\\

\midrule

Post VQC, observable construction heuristic
&
-- power spectral density mean\newline
-- standard deviation\newline
-- maximum\newline
-- spectral entropy
&
-- peak frequency\newline
-- power spectral density mean\newline
-- standard deviation\newline
-- minimum\newline
-- spectral entropy
&
\nth{1} order Pauli expansion circuit
&
amplitude embedding
&
--
&
single- or multi-qubit Pauli X, Y, Z
&
-- observables locality\newline
-- learning rate\newline
-- batch size\newline
-- n epochs
&
output expectation values processed by MLP
&
binary cross entropy
&
binary focal cross entropy
\\

\midrule

Post VQC, ansatz expansion heuristic
&
-- standard deviation\newline
-- maximum
&
-- standard deviation\newline
-- maximum
&
\nth{1} order Pauli expansion circuit
&
\nth{1} order Pauli expansion circuit
&
two-layer circuit with $R_Y$ gates and cyclic CNOT gates
&
first-qubit Pauli Z
&
-- expansion\newline
-- learning rate\newline
-- batch size\newline
-- n epochs
&
output expectation values processed by MLP
&
binary cross entropy
&
binary focal cross entropy
\\

\midrule

Post VQC, observable construction heuristic +
ansatz expansion heuristic
&
-- peak frequency\newline
-- power spectral density mean\newline
-- standard deviation\newline
-- mean\newline
-- minimum\newline
-- maximum\newline
-- quantile 0.25\newline
-- quantile 0.50\newline
-- quantile 0.75\newline
-- spectral entropy
&
--
&
amplitude embedding
&
--
&
two-layer circuit with $R_Y$ rotations and cyclic CNOT gate
&
first-qubit Pauli Z
&
-- locality\newline
-- expansion\newline
-- learning rate\newline
-- batch size\newline
-- n epochs
&
output expectation values processed by MLP
&
binary cross entropy
&
binary focal cross entropy
\\

\midrule

Data reuploading Classifier
&
-- standard deviation\newline
-- mean
&
-- power spectral density mean
&
angle embedding with rotation
&
angle embedding with rotation
&
--
&
--
&
--
&
hermitian operator
&
custom (Eq.~\ref{eq:datareup_cost})
&
custom (Eq.~\ref{eq:datareup_cost})
\\

\midrule

QLSTM
&
-- peak frequency\newline
-- power spectral density mean\newline
-- standard deviation\newline
-- quantile 0.25\newline
-- quantile 0.50\newline
-- spectral entropy
&
-- peak frequency\newline
-- power spectral density mean\newline
-- standard deviation\newline
-- maximum\newline
-- spectral entropy
&
angle embedding with $R_Y$ gates
&
angle embedding with $R_Y$ gates
&
two-layer circuit with arbitrary rotations and cyclic CNOT gates
&
multi-qubit Pauli Z
&
-- n qubits\newline
-- learning rate\newline
-- batch size\newline
-- n layers\newline
-- n epochs
&
classical LSTM workflow
&
cross entropy
&
cross entropy
\\

\midrule

Quantum Kernel SVM
&
-- power spectral density mean\newline
-- standard deviation\newline
-- quantile 0.25\newline
-- quantile 0.50
&
--
&
QAOA-inspired embedding
&
--
&
--
&
pairwise state overlaps
&
-- C (SVM regularisation)
&
classical SVM workflow
&
distance
&
--
\\

\bottomrule
\end{tabular}

\caption{Overview of the quantum models used to identify GW signals and predict the coalescence time $t_c$ in the Sangria blind dataset.}
\label{tab:quantum_models}
\end{sidewaystable}

\section{Classical Models} \label{sec:classicalmodels}

For benchmarking, we employ several standard classical models with architectures analogous to the quantum ones, namely SVM \cite{SVM_original_paper}, MLP \cite{MLP_original_paper}, RNN \cite{RNN_original_paper}, and LSTM \cite{LSTM_original_paper}. These models are implemented using standard Scikit-learn \cite{scikit-learn} and Keras \cite{Keras} implementations without architectural modifications. Their configurations and hyperparameters are summarized in Table~\ref{tab:classical_models}.

\begin{sidewaystable}[htbp]
\centering
\scriptsize
\setlength{\tabcolsep}{4pt}
\renewcommand{\arraystretch}{1.1}

\begin{tabular}{p{3.0cm}
p{2.8cm}
p{2.8cm}
p{3.2cm}
p{3.2cm}
p{2.5cm}
p{1.8cm}
p{1.8cm}}
\toprule

\multirow{2}{*}{\textbf{Model}}
& \multicolumn{2}{c}{\textbf{GW signal features}}
& \multicolumn{2}{c}{\textbf{Architecture}}
& \multirow{2}{*}{\textbf{Hyperparameters}}
& \multicolumn{2}{c}{\textbf{Loss}} \\

\cmidrule(lr){2-3}
\cmidrule(lr){4-5}
\cmidrule(lr){7-8}

& \textbf{GW detection}
& \textbf{Merger detection}
& \textbf{GW detection}
& \textbf{Merger detection}
&
& \textbf{GW detection}
& \textbf{Merger detection}
\\

\midrule

SVM
&
-- peak frequency\newline
-- power spectral density mean\newline
-- standard deviation\newline
-- maximum\newline
-- quantile 0.25\newline
-- quantile 0.50\newline
-- spectral entropy
&
-- peak frequency\newline
-- power spectral density mean\newline
-- minimum\newline
-- spectral entropy
&
sigmoid kernel
&
--
&
C
&
distance
&
--
\\

\midrule

MLP
&
-- peak frequency\newline
-- power spectral density mean\newline
-- standard deviation\newline
-- minimum\newline
-- quantile 0.25\newline
-- quantile 0.50\newline
-- quantile 0.75\newline
-- spectral entropy
&
-- peak frequency\newline
-- power spectral density mean\newline
-- minimum\newline
-- spectral entropy
&
-- Dense layer with 4 neurons and ReLu activation\newline
-- Dense layer with 32 neurons and Sigmoid activation\newline
-- Dense layer with 2 neurons and Softmax activation
&
-- Dense layer with 4 neurons and ReLu activation\newline
-- Dense layer with 8 neurons and Sigmoid activation\newline
-- Dense layer with 1 neuron and Sigmoid activation
&
batch size, learning rate, epochs
&
sparse categorical crossentropy
&
binary focal cross entropy
\\

\midrule

RNN
&
-- peak frequency\newline
-- power spectral density mean\newline
-- standard deviation\newline
-- maximum\newline
-- quantile 0.25\newline
-- quantile 0.50\newline
-- quantile 0.75\newline
-- spectral entropy
&
-- peak frequency\newline
-- power spectral density mean\newline
-- minimum\newline
-- spectral entropy
&
-- SimpleRNN layer with 10 units and ReLu activation\newline
-- Dense layer with 2 neurons and Softmax activation
&
-- SimpleRNN layer with 5 units and ReLu activation\newline
-- Dense layer with 1 neuron and Sigmoid activation
&
batch size, learning rate, epochs
&
sparse categorical crossentropy
&
binary focal cross entropy
\\

\midrule

LSTM NN
&
-- peak frequency\newline
-- power spectral density mean\newline
-- standard deviation\newline
-- maximum\newline
-- quantile 0.25\newline
-- quantile 0.50\newline
-- quantile 0.75\newline
-- spectral entropy
&
-- peak frequency\newline
-- standard deviation\newline
-- minimum\newline
-- spectral entropy
&
-- LSTM layer with 5 units\newline
-- Dense layer with 1 neuron and Sigmoid activation
&
-- LSTM layer with 5 units\newline
-- Dense layer with 1 neuron and Sigmoid activation
&
batch size, learning rate, epochs
&
binary crossentropy
&
binary focal cross entropy
\\

\bottomrule
\end{tabular}

\caption{Overview of the classical models used to identify GW signals and predict the coalescence time $t_c$ in the Sangria blind dataset.}
\label{tab:classical_models}
\end{sidewaystable}

\section{Results}

In this section, we present the results obtained by applying the quantum and classical models described in Secs. \ref{sec:quantummodels} and \ref{sec:classicalmodels} to the detection of signals and the prediction of the coalescence time, $t_c$, in the Sangria blind dataset. We compare the performance of the quantum and classical models and assess whether the results provide evidence of a quantum advantage.

\subsection{GW Signal Detection}

We first assess the ability of the quantum and classical models to identify GW signals in the Sangria blind dataset, i.e. classifying all the samples derived from the Sangria blind dataset as GW signal. The comparison considers their learning behaviour, decision regions, confusion matrices, accuracy, and F1 score.

\begin{figure}[htbp]

    \centering
    \begin{subfigure}[t]{0.48\textwidth}
        \centering
        \includegraphics[width=\linewidth]{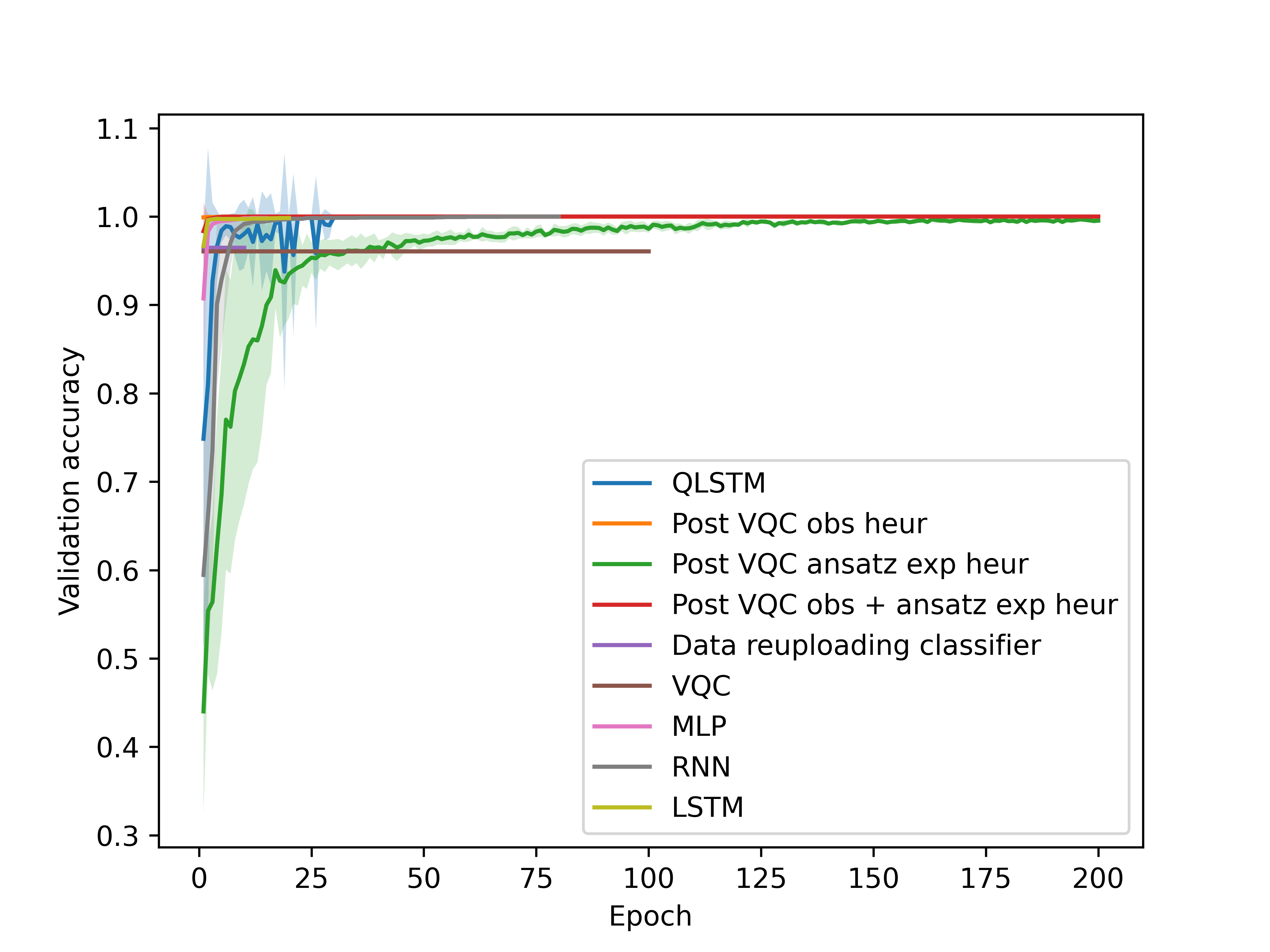}
        \caption{}
        \label{fig:learning_curves_all_val}
    \end{subfigure}
    \hfill
    \begin{subfigure}[t]{0.48\textwidth}
        \centering
        \includegraphics[width=\linewidth]{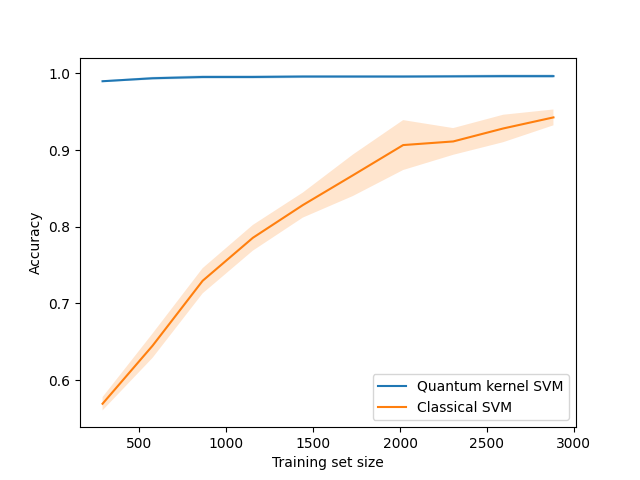}
        \caption{}
        \label{fig:learning_curves_all_val_SVM}
    \end{subfigure}

    \caption{Validation learning curves for the quantum and classical models. (a) Validation accuracy learning curves for the quantum and classical models as a function of the training epoch. The shaded regions represent standard deviation across repeated runs. (b) Validation accuracy of the quantum kernel and classical SVMs as a function of the training dataset size. The shaded regions represent the standard deviation across repeated runs.}

\end{figure}

Figure~\ref{fig:learning_curves_all_val} shows the evolution of the validation accuracy for the quantum and classical models. Among the classical models, the MLP and LSTM converge rapidly, reaching a validation accuracy close to unity within the first few epochs. The RNN starts from a lower validation accuracy and converges more gradually, but attains comparable performance after approximately 10 epochs. A similar variation in convergence behaviour is observed among the quantum models. The post VQC model based on the observables heuristic and the post VQC model combining the observables and ansatz expansion heuristics reach a validation accuracy close to unity almost immediately and remain stable throughout training. The QLSTM also converges rapidly, although its validation accuracy exhibits noticeable fluctuations during approximately the first 25 epochs. The post VQC model employing the ansatz expansion heuristic displays the slowest convergence, increasing gradually from approximately $0.44$ to a validation accuracy close to unity after approximately 150 epochs. By contrast, the standard VQC rapidly reaches a validation accuracy of approximately $0.96$, but subsequently shows no further improvement. Similarly, the data reuploading classifier maintains an approximately constant validation accuracy of $0.966$ throughout the 10 training epochs, showing no appreciable improvement with further training. Overall, most quantum and classical models achieve validation accuracies close to unity, although their convergence rates differ substantially. The classical MLP and LSTM and the two post VQC configurations using the observables heuristic exhibit the fastest convergence. The standard VQC and the data reuploading classifier attain lower validation accuracies than the other models and show little or no improvement after their initial epochs. Therefore, the validation learning curves do not indicate a general advantage of the quantum models over the classical models, although the post VQC configurations employing the observables heuristic perform comparably to the best classical models.

Figure~\ref{fig:learning_curves_all_val_SVM} shows the validation accuracy of the quantum kernel and classical SVMs as a function of the training dataset size. The quantum kernel SVM achieves an accuracy of approximately $0.99$ even for the smallest training set, after which its performance remains nearly constant as the number of training samples increases. By contrast, the classical SVM exhibits a strong dependence on the training dataset size: its accuracy increases from approximately $0.57$ for the smallest training dataset to approximately $0.94$ for the largest one. The uncertainty associated with the classical SVM is also larger, particularly for intermediate training-set sizes, whereas the quantum kernel SVM displays only limited variability. Thus, within the range of training dataset sizes considered here, the quantum kernel SVM consistently outperforms the classical SVM and appears to require fewer training samples to achieve a high validation accuracy.

\begin{figure}[htbp]
    \centering
    \setlength{\tabcolsep}{2pt}   
    \renewcommand{\arraystretch}{0.2}
    \begin{tabular}{ccc}
        \includegraphics[width=0.32\linewidth]{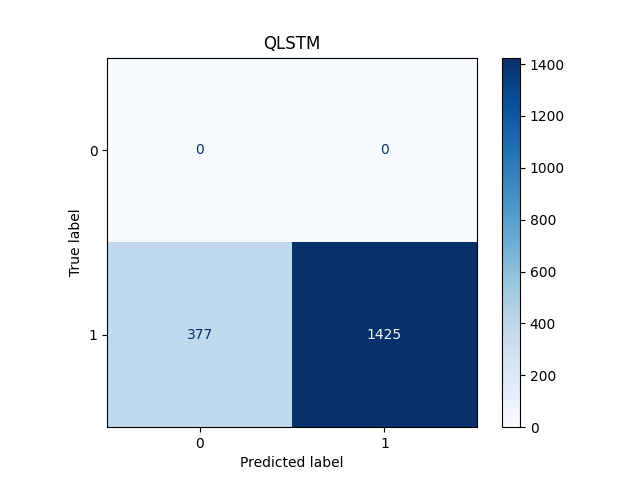} &
        \includegraphics[width=0.32\linewidth]{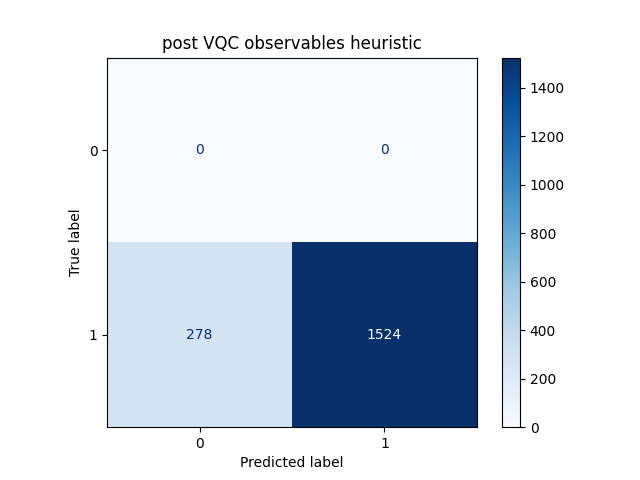} &
        \includegraphics[width=0.32\linewidth]{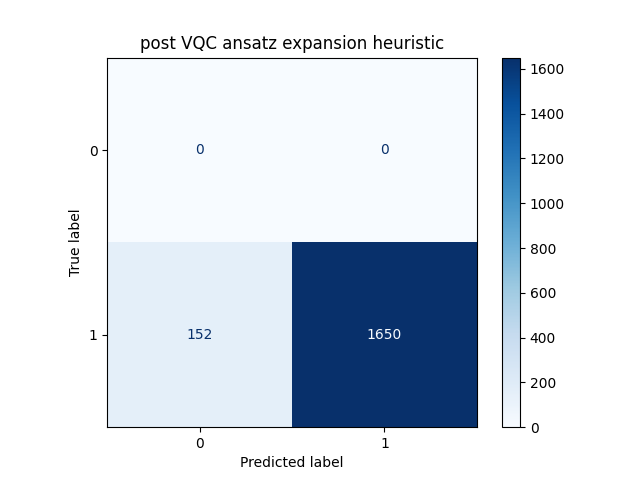} \\

        \includegraphics[width=0.32\linewidth]{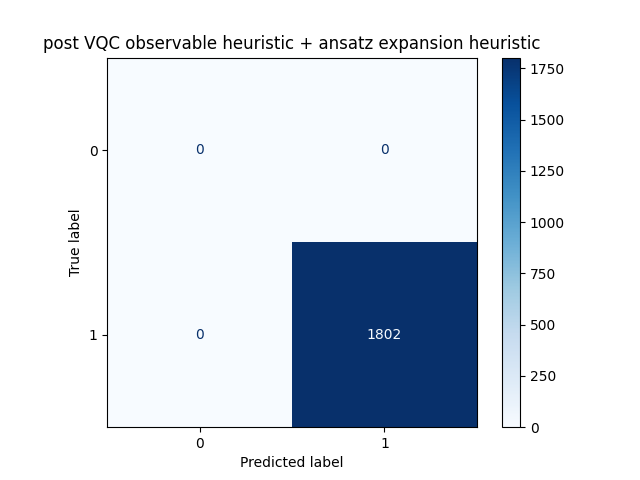} &
        \includegraphics[width=0.32\linewidth]{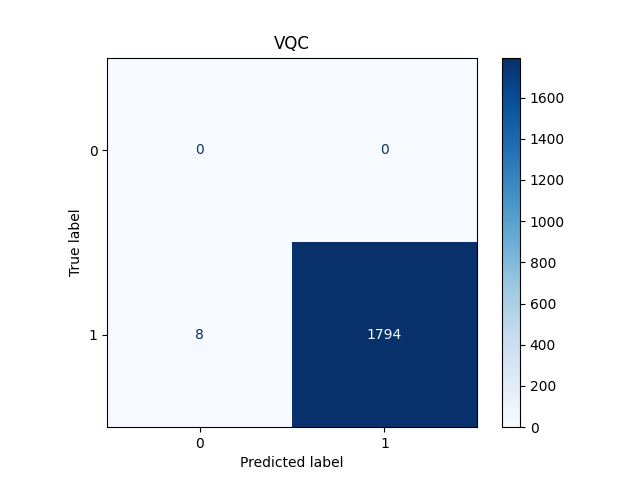} &
        \includegraphics[width=0.32\linewidth]{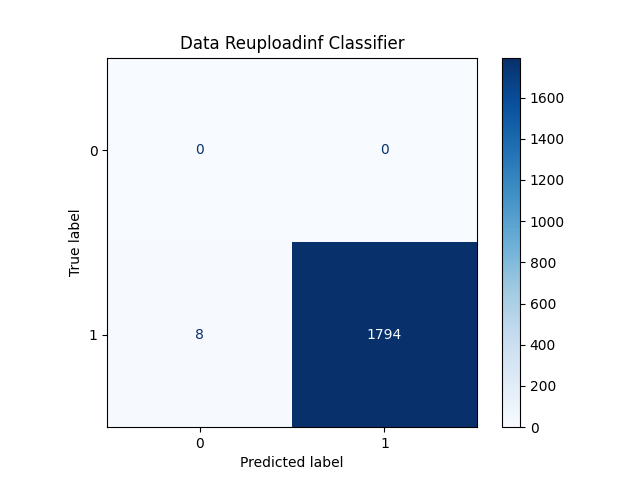} \\

        \includegraphics[width=0.32\linewidth]{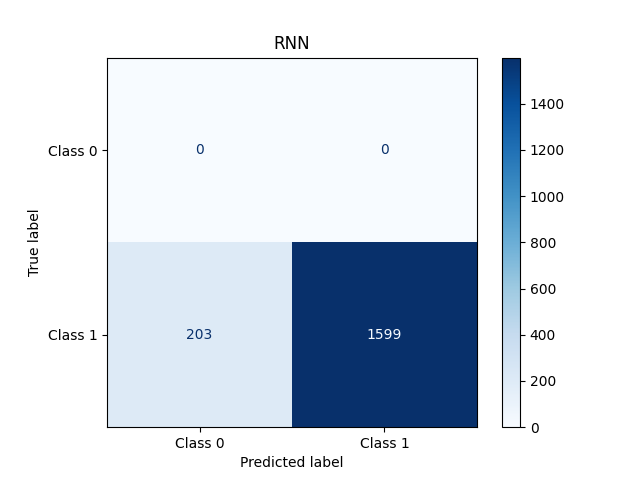} &
        \includegraphics[width=0.32\linewidth]{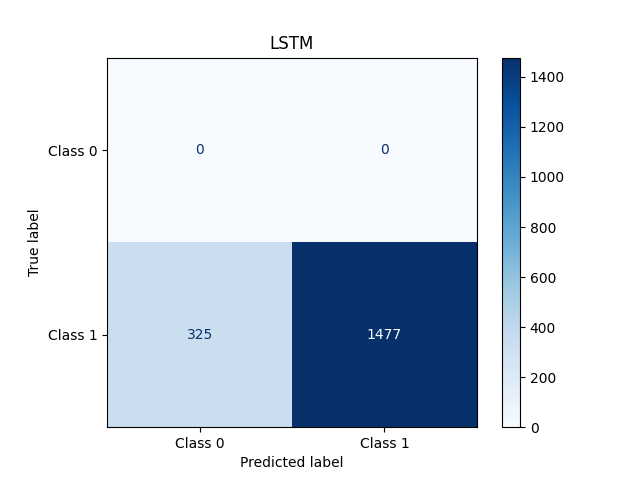} &
        \includegraphics[width=0.32\linewidth]{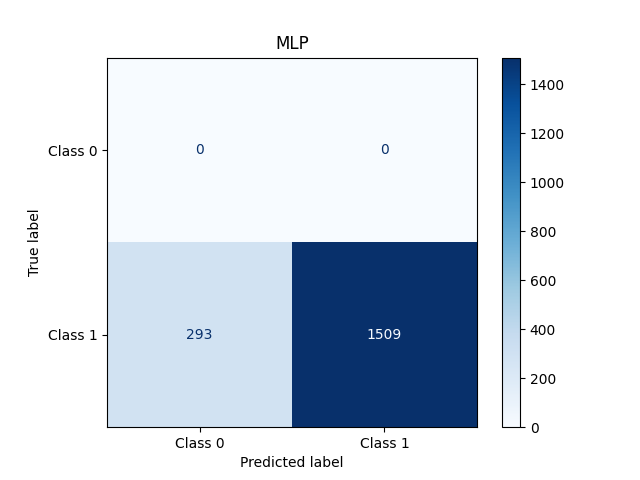} \\

        \includegraphics[width=0.32\linewidth]{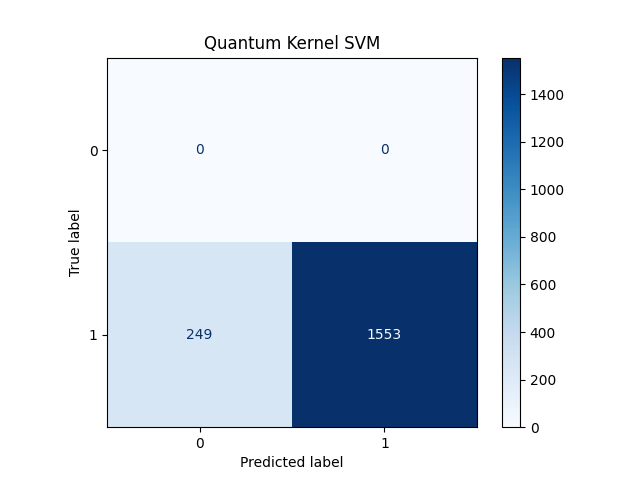} &
        \includegraphics[width=0.32\linewidth]{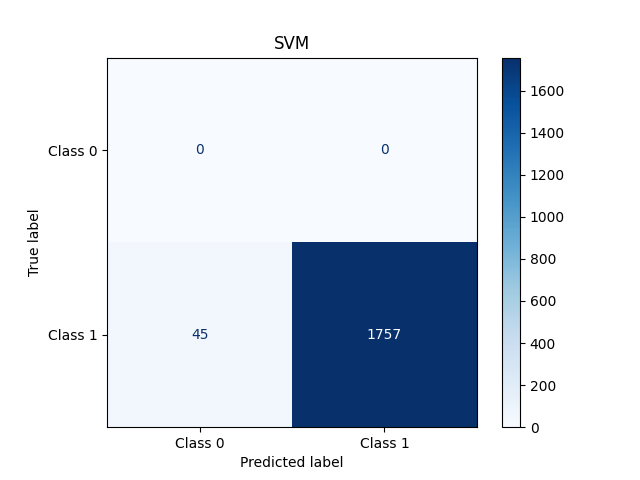} &
    \end{tabular}
    \caption{Confusion matrices for quantum and classical models.}
    \label{fig:confusion_matrices_all}
\end{figure}

Figure~\ref{fig:confusion_matrices_all} shows the confusion matrices obtained for the quantum and classical models on the blind dataset, which contains only GW signal (Class 1) samples. The post VQC model combining the observables and ansatz expansion heuristics correctly classifies all 1802 samples. The VQC and the data reuploading classifier also achieve a high detection rate, each correctly classifying 1794 samples and misclassifying 8 samples as noise (Class 0). The post VQC model using only the ansatz expansion heuristic correctly identifies 1650 samples, while the model using only the observables heuristic correctly identifies 1524 samples. The QLSTM correctly classifies 1425 samples and misclassifies 377 samples, corresponding to the largest number of missed detections among the considered quantum models. Among the classical models, the classical SVM correctly identifies 1757 samples and misclassifies 45 samples, achieving the best performance within this group. The RNN correctly classifies 1599 samples, followed by the MLP and LSTM, which correctly classify 1509 and 1477 samples, respectively. The quantum kernel SVM correctly identifies 1553 samples and misclassifies 249 samples, and is therefore outperformed by the classical SVM on this prediction dataset. Overall, the post VQC model combining the two heuristics produces the fewest missed detections, followed by the VQC and the data reuploading classifier, outperforming classical models.

\begin{figure}[htbp]
    \centering
    \begin{subfigure}{0.32\linewidth}
        \centering
        \includegraphics[width=\linewidth]{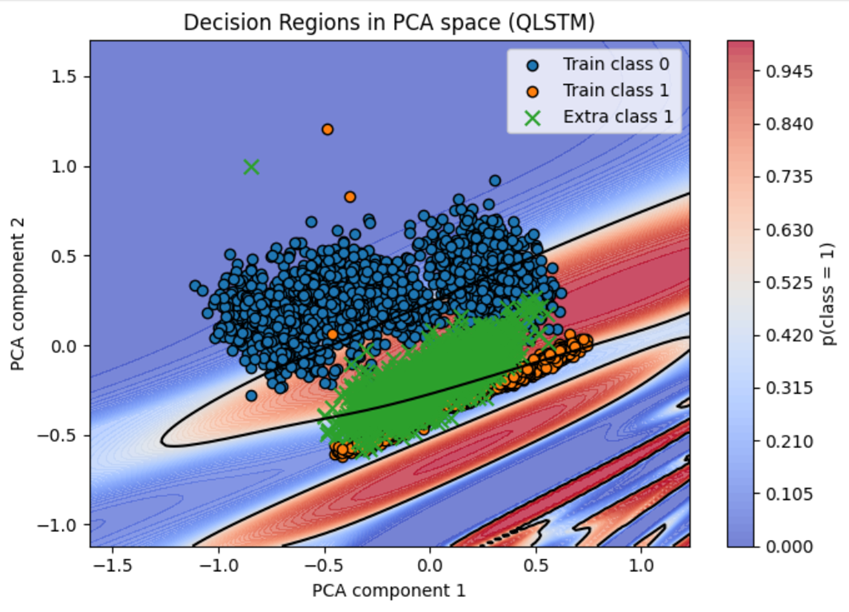}
        \caption{}
    \end{subfigure}
    \begin{subfigure}{0.32\linewidth}
        \centering
        \includegraphics[width=\linewidth]{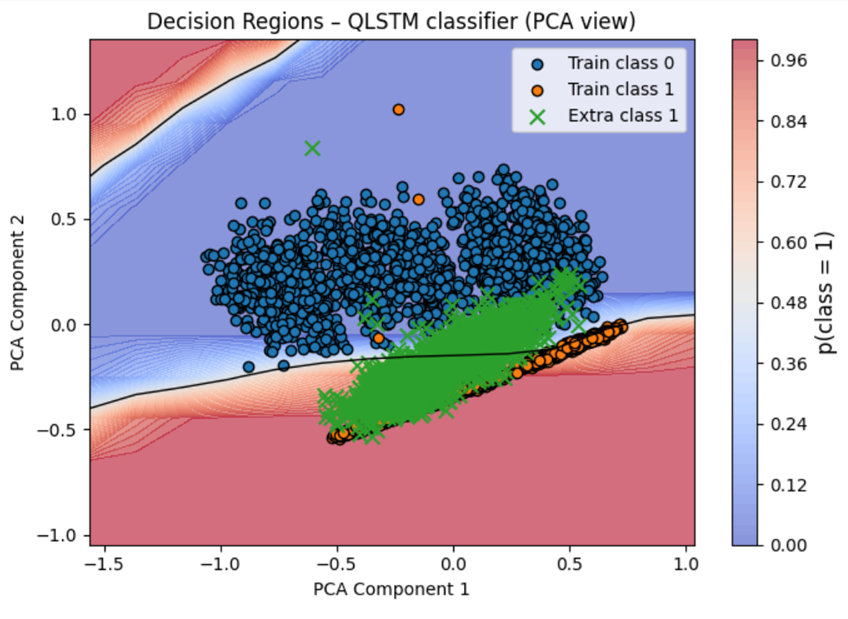}
        \caption{}
    \end{subfigure}
    \begin{subfigure}{0.32\linewidth}
        \centering
        \includegraphics[width=\linewidth]{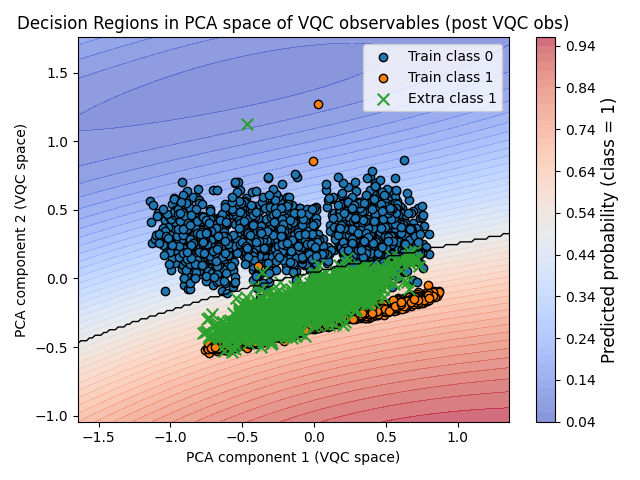}
        \caption{}
    \end{subfigure}
    \caption{Decision regions in the PCA space for selected models.}
    \label{fig:decision_regions_PCA}
\end{figure}

\begin{figure}[htbp]
    \centering
    \setlength{\tabcolsep}{3pt}    
    \renewcommand{\arraystretch}{0.9}
    \begin{tabular}{ccc}
        \includegraphics[width=0.32\linewidth]{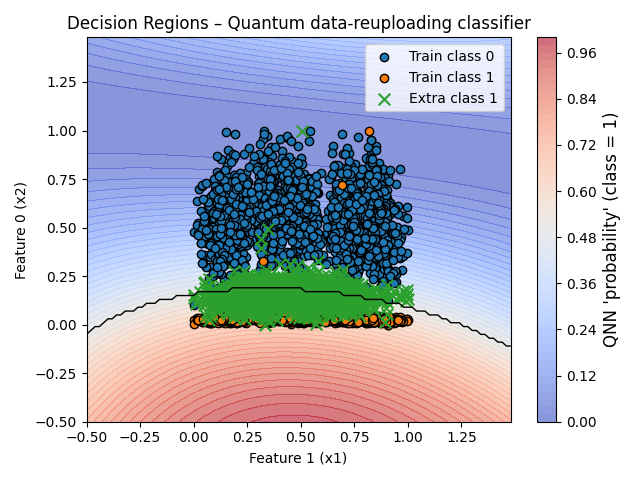} &
        \includegraphics[width=0.32\linewidth]{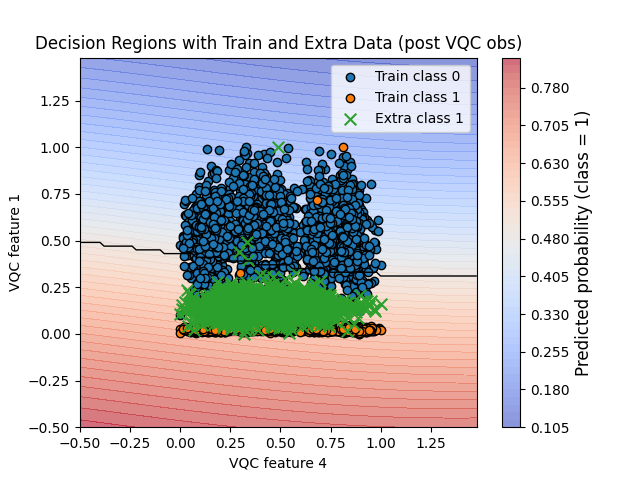} &
        \includegraphics[width=0.32\linewidth]{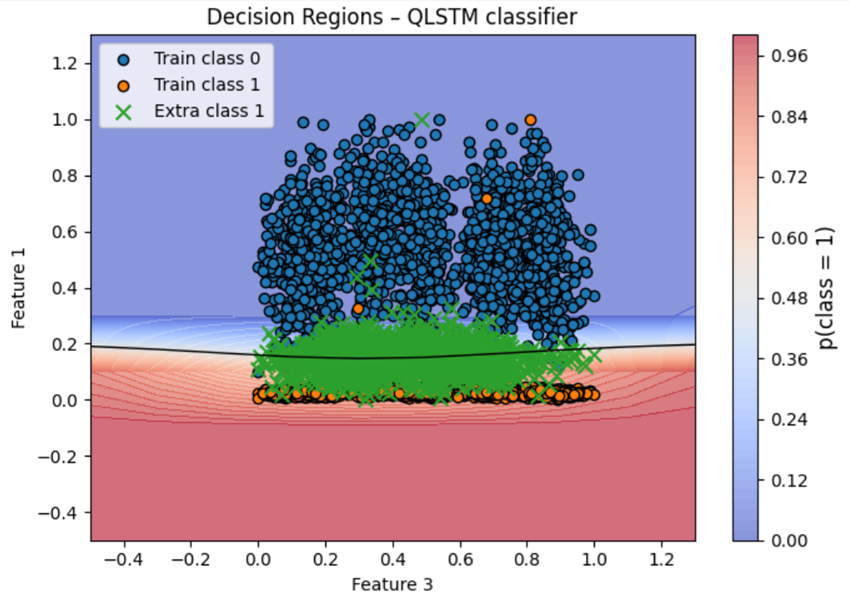} \\

        \includegraphics[width=0.32\linewidth]{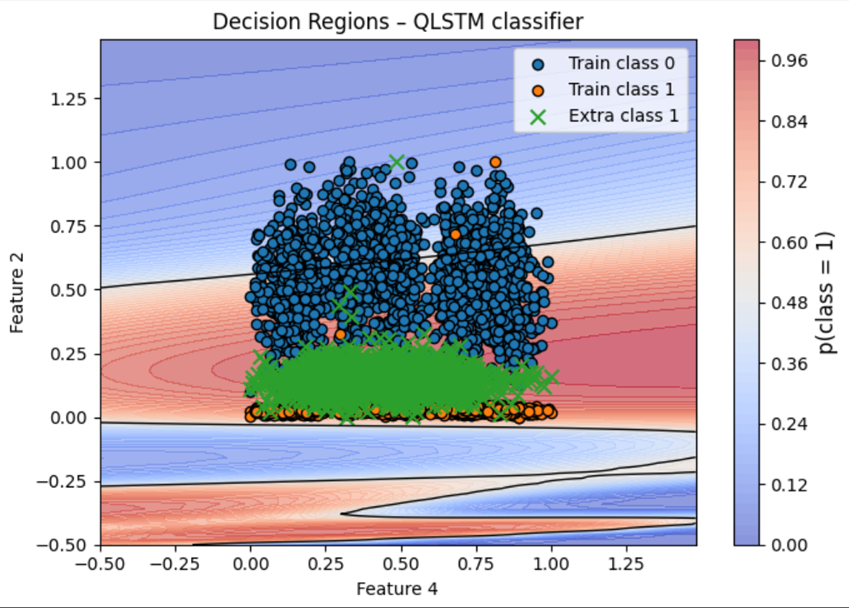} &
        \includegraphics[width=0.32\linewidth]{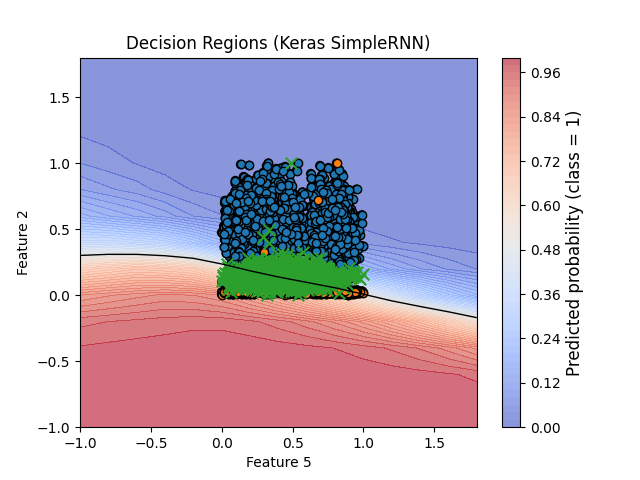} &
        \includegraphics[width=0.32\linewidth]{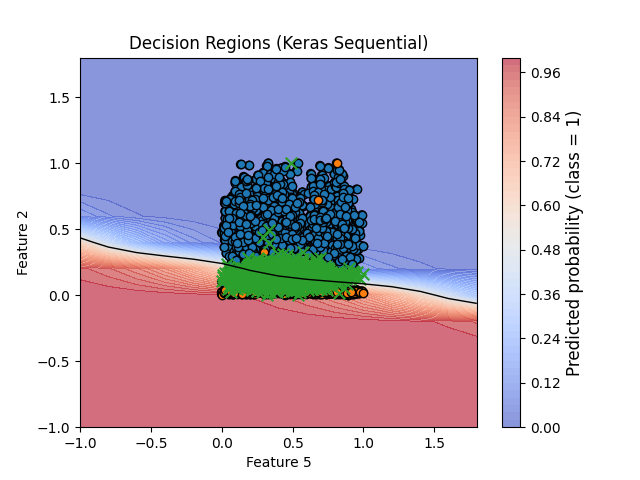} \\

        \includegraphics[width=0.32\linewidth]{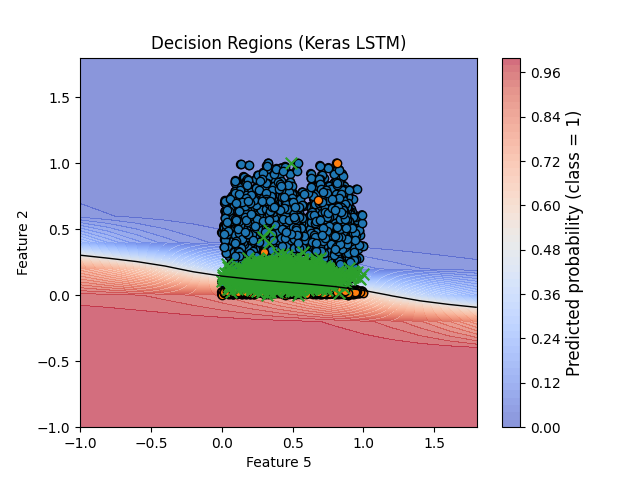} &
        \includegraphics[width=0.32\linewidth]{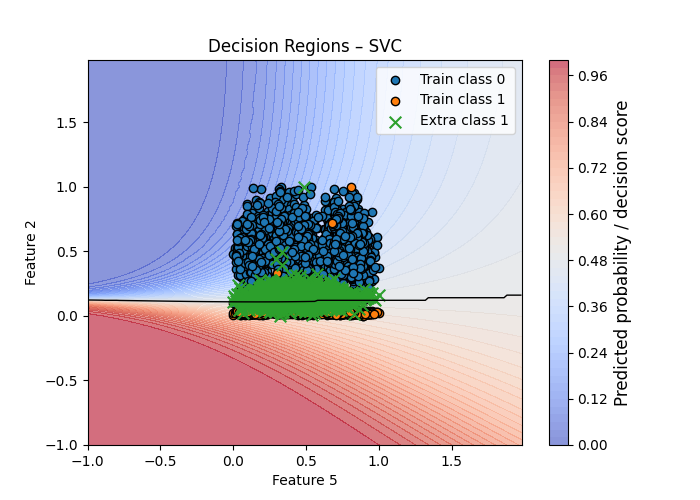} &
        {} 
    \end{tabular}
    \caption{Decision regions for selected models in the standard deviation (Feature 2) VS maximum (feature 5) space.}
    \label{fig:decision_regions_optim}
\end{figure}

Figure \ref{fig:decision_regions_optim} and Figure \ref{fig:decision_regions_PCA} show the decision regions obtained for the quantum and classical models. The training samples occupy two partially separated regions of the feature space: the Class 0 samples are concentrated predominantly at larger values of the standard deviation feature, whereas the Class 1 samples are located closer to the lower part of the distribution. Accordingly, most models learn a boundary that separates the compact Class 1 region from the broader Class 0 distribution. The additional Class 1 samples are generally concentrated on or below this boundary, although some samples extend into regions assigned to Class 0. These points correspond to the Class 1 samples that are more difficult for the models to identify. The decision regions produced by the models differ substantially in their complexity. The VQC based on the observables heuristic, the post VQC models, the data reuploading classifier, and most classical models produce comparatively smooth decision boundaries that broadly follow the separation between the two training classes. The RNN, MLP, LSTM, and SVC boundaries are predominantly smooth and approximately linear or weakly curved over the region occupied by the data. The data reuploading classifier produces a more noticeably curved boundary, adapting to the shape of the Class 1 distribution. By contrast, some QLSTM and VQC configurations produce more complex decision regions, including multiple narrow or disconnected areas with different predicted probabilities. In particular, the striped regions visible outside the main concentration of training samples indicate substantial variation in the model output in sparsely populated parts of the projected feature space. However, these structures occur mainly outside the region containing most of the samples and therefore do not necessarily imply improved separation of the two classes. The distribution of the additional Class 1 samples relative to the decision boundaries is consistent with the confusion matrix results (Fig. \ref{fig:confusion_matrices_all}). Models for which most green points lie within the Class 1 decision region produce fewer missed detections, whereas models whose boundaries intersect a larger portion of the green distribution classify more GW signals as noise. In particular, the position and shape of the boundary near the overlap between the training classes appear to determine how many additional Class 1 samples are correctly detected. Since the decision regions are visualised in two-dimensional projections or selected pairs of features, they provide a qualitative representation of the classifiers and may not fully describe the decision functions in the complete feature space.

Tables~\ref{tab:quantum_results_pre_sangria} and \ref{tab:classical_results_pre_sangria} show that most quantum and classical models achieve training and validation accuracies close to unity. However, several models exhibit a substantial decrease in performance on the blind (text) dataset. Since this dataset contains only Class 1 samples, the Class 1 recall is used to quantify the fraction of GW signals correctly detected. Among the quantum models, the post VQC combining the observables and ansatz expansion heuristics achieves the highest recall, $\mathrm{Recall}=1.00$, correctly detecting all GW signals. The VQC and the data reuploading classifier follow, both obtaining a recall of $0.996$. The post VQC employing only the ansatz expansion heuristic achieves a recall of $0.916$, while the model employing only the observables heuristic obtains a recall of $0.846$. The QLSTM exhibits the largest decrease between validation accuracy and performance on the blind dataset, with its validation accuracy of $1.00$ corresponding to a recall of only $0.791$ on the blind dataset. Among the classical models, the SVM achieves the highest recall, $\mathrm{Recall}=0.975$. Although the MLP, RNN, and LSTM reach validation accuracies of $1.00$, their recall values on the blind dataset decrease to $0.837$, $0.887$, and $0.820$, respectively. The quantum kernel SVM obtains a recall of $0.862$ and is therefore outperformed by the classical SVM. Overall, the post VQC combining both heuristics obtains the highest recall, followed by the VQC and the data reuploading classifier. However, the performance varies considerably across the quantum architectures, and the classical SVM outperforms several quantum models. Consequently, these results indicate an advantage for specific quantum configurations rather than a general quantum advantage. Moreover, because the blind dataset contains only Class 1 samples, this comparison characterises the models' ability to detect GW signals but does not assess their ability to identify noise samples.

\begin{table}[htbp]
\centering
\scriptsize
\setlength{\tabcolsep}{0pt}   
\renewcommand{\arraystretch}{1.15}

\begin{tabular}{@{}
p{8.0 cm} 
C{2.0 cm} 
C{2.0 cm} 
C{2.0 cm}
C{2.0 cm} 
@{}}
\toprule
\textbf{Model} &
$\mathbf{ACC}_{\text{train}}$ &
$\mathbf{ACC}_{\text{val}}$ &
$\mathbf{ACC}_{\text{test}}$ &
$\mathbf{F1}$ \\
\midrule

VQC
& 0.95 & 0.96 & 0.99 & 0.91 \\
\midrule

Post VQC, observable construction heuristic
& 0.99 & 0.99 & 0.90 & 0.91 \\
\midrule

Post VQC, ansatz expansion heuristic
& 0.99 & 0.99 & 0.95 & 0.95 \\
\midrule

Post VQC, observable construction heuristic + ansatz expansion heuristic
& 0.99 & 1.00 & 1.00 & 0.91 \\
\midrule

Data Reuploading Classifier
& 0.98 & 0.98 & 0.98 & 0.9 \\
\midrule

QLSTM
& 0.99 & 1.00 & 0.79 & 0.88 \\
\midrule

Quantum Kernel SVM
& -- & 0.99 & 0.86 & 0.92 \\

\bottomrule
\end{tabular}

\caption{Training, validation, and test accuracies and F1 scores obtained by the quantum models before their application to the Sangria dataset.}
\label{tab:quantum_results_pre_sangria}
\end{table}

\begin{table}[htbp]
\centering
\scriptsize
\setlength{\tabcolsep}{0pt}  
\renewcommand{\arraystretch}{1.15}

\begin{tabular}{@{}
p{8.0 cm}  
C{2.0 cm}  
C{2.0 cm}  
C{2.0 cm}  
C{2.0 cm} 
@{}}
\toprule
\textbf{Model} &
$\mathbf{ACC}_{\text{train}}$ &
$\mathbf{ACC}_{\text{val}}$ &
$\mathbf{ACC}_{\text{test}}$ &
$\mathbf{F1}$ \\
\midrule

SVM
& 0.95 & -- & 0.97 & 0.98 \\
\midrule

MLP
& 0.99 & 1 & 0.83 & 0.91 \\
\midrule

RNN
& 0.99 & 1.00 & 0.88 & 0.94 \\
\midrule

LSTM NN
& 0.99 & 1.00 & 0.82 & 0.90 \\

\bottomrule
\end{tabular}

\caption{Training, validation, and test accuracies and F1 scores obtained by the classical models before their application to the Sangria blind dataset.}
\label{tab:classical_results_pre_sangria}
\end{table}

The best performing quantum models combine high validation accuracy with strong generalisation to the blind dataset. In particular, the post VQC combining both heuristics correctly identifies all GW signals, while the standard VQC and data reuploading classifier also outperform all classical models. However, this advantage is not consistent across quantum architectures: the classical SVM outperforms the QLSTM, quantum-kernel SVM, and some post VQC configurations. Moreover, the greater complexity of some quantum decision regions does not systematically translate into better detection performance. Overall, the results indicate an advantage for specific quantum configurations rather than a general quantum advantage.

\subsection{Merger Detection and Coalescence Time Estimation}

We next evaluate the ability of the quantum and classical models to predict the merger time, $t_c$, for the detected signals. The merger prediction performance is evaluated by determining whether the true merger time, $t_c$, lies within the interval predicted by the model and by measuring how far in advance of the true $t_c$ the merger is detected, allowing us to assess the accuracy and robustness of each model.

\begin{figure}
    \centering
    \begin{subfigure}[b]{\linewidth}
        \centering
        \includegraphics[width=\linewidth]{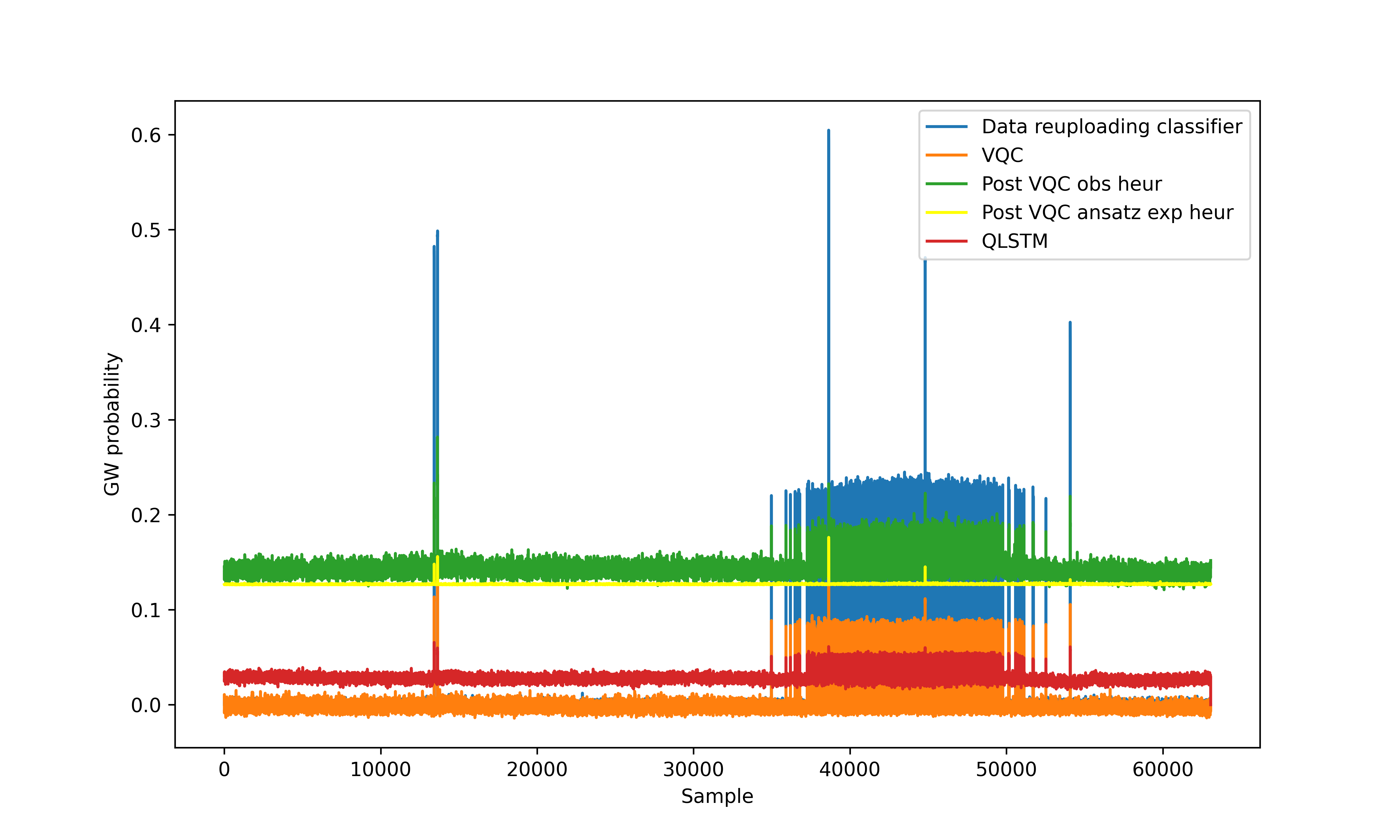}
        \caption{}
        \label{fig:gw_probs_all_new_final}
    \end{subfigure}

    \vspace{1em}

    \begin{subfigure}[b]{\linewidth}
        \centering
        \includegraphics[width=\linewidth]{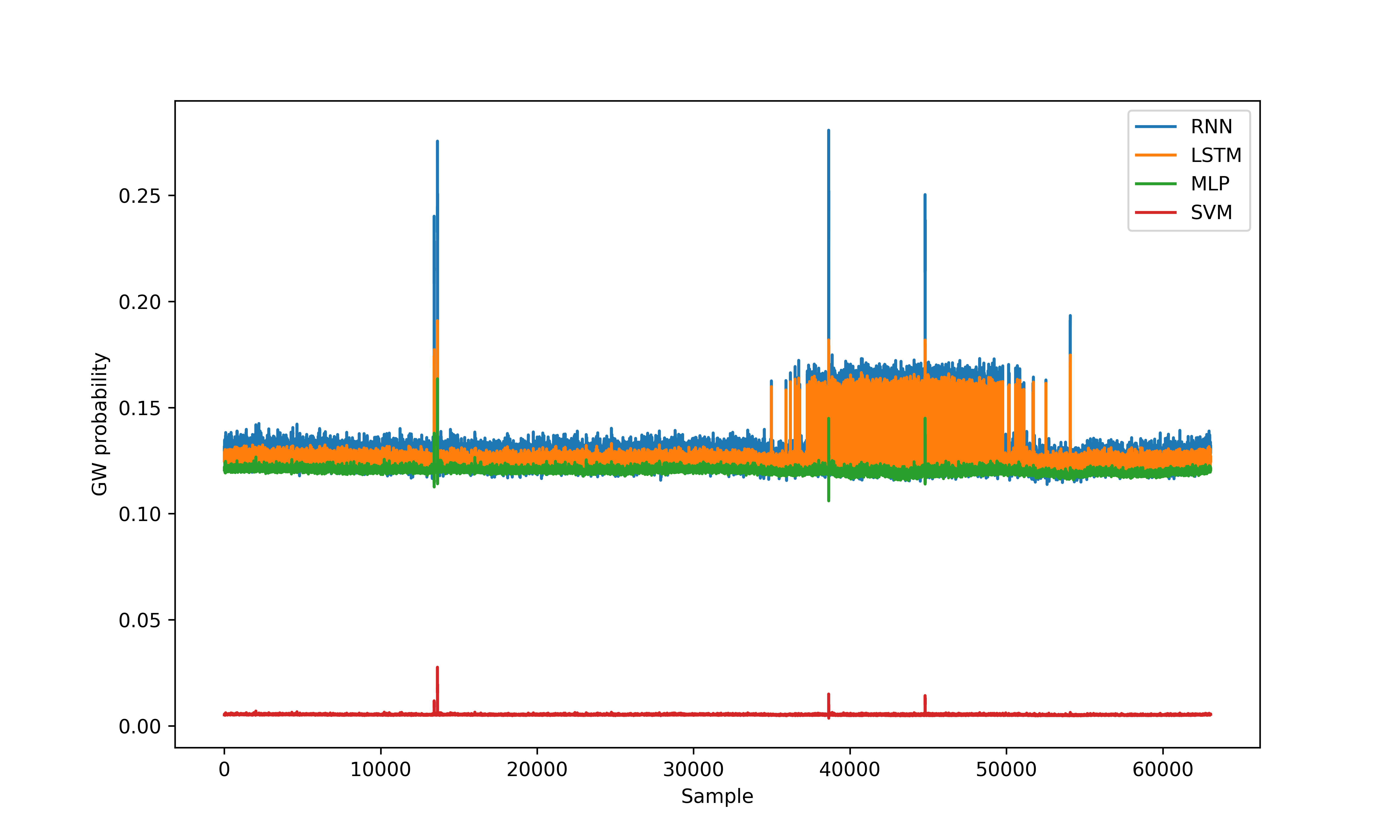}
        \caption{}
        \label{fig:gw_probs_all_new_1_NNs}
    \end{subfigure}

    \caption{Probability time series produced by (a) the quantum models and (b) the classical models for the analysis of the Sangria blind dataset. Each point corresponds to the probability assigned to a moving data window. Probability thresholds are used to identify candidate merger regions.}
    \label{fig:gw_probs_all}
\end{figure}

The output of each model is a probability series (Fig.\ref{fig:gw_probs_all_new_final} and Fig.\ref{fig:gw_probs_all_new_1_NNs}), where each value corresponds to a data window. Windows containing merger events are identified by applying a heuristic model-specific threshold. For each masked probability series, the Hilbert envelope is computed and its local maxima are identified. The window containing a maximum is interpreted as the predicted coalescence time interval. The window indices are converted into physical time intervals by reversing Eq. (\ref{eq:noofmovingwindows}).

 or a window peak index $k_\text{peak}$, the predicted interval is $\left[L,U\right]$, where:

\begin{equation}
    L = k_\text{peak} N_s \Delta t
\end{equation}

\begin{equation}
    U = L + N_w \Delta t
\end{equation}

where $N_w$ is the window length, $N_s$ is the step, and $\Delta t$ is the sampling time cadence. The predicted time to merger is evaluated from the interval between the beginning of the corresponding mask and the peak position. By denoting $k_\text{start}$ denotes the first window in the mask, then the time to merger $\tau$ is:

\begin{equation}
    \tau = \left( k_\text{peak} - k_\text{start} \right) N_s \Delta t
\end{equation}

The predicted coalescence interval $\left[L,U\right]$ is then compared with the true coalescence time $t_c$. It's midpoint is 

\begin{equation}
    m = \frac{L+U}{2}
    \label{eq:midpoint}
\end{equation}

and the absolute midpoint error is

\begin{equation}
    \epsilon_{\mathrm{mid}} = \left|m-t_c\right| = \left|\frac{L+U}{2}-t_c\right|
    \label{eq:abs_midpoint_err}
\end{equation}

To quantify predictions for which the true coalescence time lies outside the interval, the outside error is defined as

\begin{equation}
    \epsilon_{\mathrm{out}} = \max\left(L-t_c,\,0,\,t_c-U\right)
    \label{eq:outside_err}
\end{equation}

Thus, $\epsilon_{\mathrm{out}} = 0$ when $t_c \in \left[L,U\right]$. The resulting detection outcomes, predicted time to merger, coalescence intervals, and localisation error are reported in Tab.\ref{tab:coalescence-interval-errors} and Tab.\ref{tab:classical_results_sangria}.

The merger detection results presented in Tables \ref{tab:coalescence-interval-errors} and \ref{tab:classical_results_sangria} reveal clear differences between the quantum and classical models in terms of both detection capability and coalescence time localisation accuracy. Overall, all models struggled to detect Signal 5, which remained undetected across both the quantum and classical approaches. Excluding this signal, the baseline VQC achieved the most consistent performance among the quantum models, successfully detecting five out of six signals and, importantly, producing predicted coalescence intervals that always contained the true merger time $t_c$. This is reflected by the zero outside error obtained for all detected events. Furthermore, the VQC exhibits the smallest absolute midpoint errors among all investigated models, ranging from 249~s to 1410~s, indicating accurate localisation of the coalescence event.

The post VQC models based on observable and ansatz expansion heuristics maintained comparable detection rates but generally exhibited larger localisation errors. The observable heuristic correctly enclosed the true $t_c$ for only two of the five detected signals, while the ansatz expansion heuristic achieved the same result for only one signal. Although these approaches occasionally provided longer advance warning times, reaching up to 1.53~h before merger, this gain was accompanied by reduced localisation accuracy. Notably, combining both heuristics resulted in a complete failure to identify any merger candidates, suggesting that the increased model complexity introduced by the combined approach adversely affected generalisation on the analysed dataset. The Data Reuploading Classifier achieved a similar detection rate to the baseline VQC but generated fewer accurate coalescence intervals. Three of the five detected mergers were localised correctly, while the remaining predictions exhibited outside errors between 905~s and 1560~s. A comparable behaviour is observed for the QLSTM model, which successfully detected five signals but correctly enclosed the true coalescence time for only two events.

\begin{table*}[htbp]
\centering
\scriptsize
\setlength{\tabcolsep}{2pt}
\renewcommand{\arraystretch}{1.15}

\begin{tabular}{@{}
p{3.9cm}
C{0.55cm}
C{0.85cm}
C{1.55cm}
C{2.15cm}
C{1.25cm}
C{1.25cm}
C{1.05cm}
C{0.9cm}
@{}}
\toprule
\textbf{Model} &
\textbf{Signal} &
\textbf{Detection} &
\textbf{Predicted time to coalescence (h)} &
\textbf{Predicted coalescence interval (s)} &
\textbf{Interval midpoint (s)} &
\textbf{Absolute midpoint error (s)} &
\textbf{Outside error (s)} &
\textbf{Inside interval} \\
\midrule

\multirow{6}{*}{%
  \parbox{3.9cm}{\raggedright VQC}
}
& 1 & \checkmark & 0.695 & 6709000--6714000 & 6711500 & 905 & 0 & \checkmark \\
\cline{2-9}
& 2 & \checkmark & 0.834 & 6816000--6821000 & 6818500 & 249 & 0 & \checkmark \\
\cline{2-9}
& 3 & \checkmark & 0.834 & 19320500--19325500 & 19323000 & 1060 & 0 & \checkmark \\
\cline{2-9}
& 4 & \checkmark & 0.278 & 22402000--22407000 & 22404500 & 1410 & 0 & \checkmark \\
\cline{2-9}
& 5 & \xmark & -- & -- & -- & -- & -- & -- \\
\cline{2-9}
& 6 & \checkmark & 0.973 & 27042000--27047000 & 27044500 & 370 & 0 & \checkmark \\
\midrule

\multirow{6}{*}{%
  \parbox{3.9cm}{\raggedright
  Post VQC, observable construction heuristic}
}
& 1 & \checkmark & 0.000 & 6706500--6711500 & 6709000 & 3405 & 905 & \xmark \\
\cline{2-9}
& 2 & \checkmark & 0.000 & 6813000--6818000 & 6815500 & 2751 & 251 & \xmark \\
\cline{2-9}
& 3 & \checkmark & 1.528 & 19323500--19328500 & 19326000 & 1940 & 0 & \checkmark \\
\cline{2-9}
& 4 & \checkmark & 1.250 & 22405000--22410000 & 22407500 & 1590 & 0 & \checkmark \\
\cline{2-9}
& 5 & \xmark & -- & -- & -- & -- & -- & -- \\
\cline{2-9}
& 6 & \checkmark & 0.000 & 27038500--27043500 & 27041000 & 3130 & 630 & \xmark \\
\midrule

\multirow{6}{*}{%
  \parbox{3.9cm}{\raggedright
  Post VQC, ansatz expansion heuristic}
}
& 1 & \checkmark & 1.389 & 6707000--6712000 & 6709500 & 2905 & 405 & \xmark \\
\cline{2-9}
& 2 & \checkmark & 1.250 & 6817500--6822500 & 6820000 & 1749 & 0 & \checkmark \\
\cline{2-9}
& 3 & \checkmark & 0.139 & 19318500--19323500 & 19321000 & 3060 & 560 & \xmark \\
\cline{2-9}
& 4 & \checkmark & 0.000 & 22400500--22405500 & 22403000 & 2910 & 410 & \xmark \\
\cline{2-9}
& 5 & \xmark & -- & -- & -- & -- & -- & -- \\
\cline{2-9}
& 6 & \checkmark & 0.000 & 27038500--27043500 & 27041000 & 3130 & 630 & \xmark \\
\midrule

\multirow{6}{*}{%
  \parbox{3.9cm}{\raggedright
  Post VQC, observable construction heuristic
  + ansatz expansion heuristic}
}
& 1 & \xmark & -- & -- & -- & -- & -- & -- \\
\cline{2-9}
& 2 & \xmark & -- & -- & -- & -- & -- & -- \\
\cline{2-9}
& 3 & \xmark & -- & -- & -- & -- & -- & -- \\
\cline{2-9}
& 4 & \xmark & -- & -- & -- & -- & -- & -- \\
\cline{2-9}
& 5 & \xmark & -- & -- & -- & -- & -- & -- \\
\cline{2-9}
& 6 & \xmark & -- & -- & -- & -- & -- & -- \\
\midrule

\multirow{6}{*}{%
  \parbox{3.9cm}{\raggedright
  Data Reuploading Classifier}
}
& 1 & \checkmark & 0.000 & 6706500--6711500 & 6709000 & 3405 & 905 & \xmark \\
\cline{2-9}
& 2 & \checkmark & 1.250 & 6817500--6822500 & 6820000 & 1749 & 0 & \checkmark \\
\cline{2-9}
& 3 & \checkmark & 0.000 & 19317500--19322500 & 19320000 & 4060 & 1560 & \xmark \\
\cline{2-9}
& 4 & \checkmark & 1.250 & 22405000--22410000 & 22407500 & 1590 & 0 & \checkmark \\
\cline{2-9}
& 5 & \xmark & -- & -- & -- & -- & -- & -- \\
\cline{2-9}
& 6 & \checkmark & 1.250 & 27043000--27048000 & 27045500 & 1370 & 0 & \checkmark \\
\midrule

\multirow{6}{*}{%
  \parbox{3.9cm}{\raggedright QLSTM NN}
}
& 1 & \checkmark & 0.000 & 6706500--6711500 & 6709000 & 3405 & 905 & \xmark \\
\cline{2-9}
& 2 & \checkmark & 0.000 & 6813000--6818000 & 6815500 & 2751 & 251 & \xmark \\
\cline{2-9}
& 3 & \checkmark & 0.000 & 19317500--19322500 & 19320000 & 4061 & 1561 & \xmark \\
\cline{2-9}
& 4 & \checkmark & 0.000 & 22405000--22410000 & 22407500 & 1590 & 0 & \checkmark \\
\cline{2-9}
& 5 & \xmark & -- & -- & -- & -- & -- & -- \\
\cline{2-9}
& 6 & \checkmark & 0.000 & 27043000--27048000 & 1590 & 1370 & 0 & \checkmark \\
\bottomrule
\end{tabular}

\caption{Predicted coalescence intervals and their comparison with
the true coalescence times. The absolute midpoint error is the absolute
difference between the midpoint of the predicted interval and the true
coalescence time. The outside error is zero when the true coalescence
time lies within the predicted interval.}
\label{tab:coalescence-interval-errors}
\end{table*}

Among the classical models, the LSTM provided the strongest overall performance. It correctly detected five signals and produced valid coalescence intervals for three events, matching the performance of the Data Reuploading Classifier. The RNN achieved a similar detection rate but correctly localised only two mergers. The MLP detected four signals, although only one predicted interval contained the true $t_c$. The SVM displayed the least robust behaviour, producing a catastrophic localisation error for Signal 1, where the model associated the event with the coalescence interval of Signal 2, resulting in an outside error exceeding $10^5$~s.

\begin{table*}[!tp]
\centering
\scriptsize
\setlength{\tabcolsep}{2pt}
\renewcommand{\arraystretch}{1.15}

\begin{tabular}{@{}
p{3.9cm}
C{0.55cm}
C{0.85cm}
C{1.55cm}
C{2.15cm}
C{1.25cm}
C{1.25cm}
C{1.05cm}
C{0.9cm}
@{}}
\toprule
\textbf{Model} &
\textbf{Signal} &
\textbf{Detection} &
\textbf{Predicted time to coalescence (h)} &
\textbf{Predicted coalescence interval (s)} &
\textbf{Interval midpoint (s)} &
\textbf{Absolute midpoint error (s)} &
\textbf{Outside error (s)} &
\textbf{Inside interval} \\
\midrule

\multirow{6}{*}{%
  \parbox{3.9cm}{\raggedright MLP}
}
& 1 & \checkmark & 0.139 & 6707000--6712000 & 6709500 & 2905 & 405 & \xmark \\
\cline{2-9}
& 2 & \checkmark & 0.139 & 6813500--6818500 & 6816000 & 2251 & 0 & \checkmark \\
\cline{2-9}
& 3 & \checkmark & 0.000 & 19318000--19323000 & 19320500 & 3560 & 1060 & \xmark \\
\cline{2-9}
& 4 & \checkmark & 0.000 & 22400500--22405500 & 22403000 & 2910 & 410 & \xmark \\
\cline{2-9}
& 5 & \xmark & -- & -- & -- & -- & -- & -- \\
\cline{2-9}
& 6 & \xmark & -- & -- & -- & -- & -- & -- \\
\midrule

\multirow{6}{*}{%
  \parbox{3.9cm}{\raggedright RNN}
}
& 1 & \checkmark & 0.000 & 6706500--6711500 & 6709000 & 3405 & 905 & \xmark \\
\cline{2-9}
& 2 & \checkmark & 0.000 & 6813000--6818000 & 6815500 & 2751 & 251 & \xmark \\
\cline{2-9}
& 3 & \checkmark & 0.139 & 19318000--19323000 & 19320500 & 3560 & 1060 & \xmark \\
\cline{2-9}
& 4 & \checkmark & 1.250 & 22405000--22410000 & 22407500 & 1590 & 0 & \checkmark \\
\cline{2-9}
& 5 & \xmark & -- & -- & -- & -- & -- & -- \\
\cline{2-9}
& 6 & \checkmark & 0.000 & 27038500--27043500 & 27041000 & 3130 & 630 & \xmark \\
\midrule

\multirow{6}{*}{%
  \parbox{3.9cm}{\raggedright LSTM}
}
& 1 & \checkmark & 0.000 & 6706500--6711500 & 6709000 & 3405 & 905 & \xmark \\
\cline{2-9}
& 2 & \checkmark & 1.250 & 6817500--6822500 & 6820000 & 1749 & 0 & \checkmark \\
\cline{2-9}
& 3 & \checkmark & 0.000 & 19317500--19322500 & 19320000 & 4060 & 1560 & \xmark \\
\cline{2-9}
& 4 & \checkmark & 1.250 & 22405000--22410000 & 22407500 & 1590 & 0 & \checkmark \\
\cline{2-9}
& 5 & \xmark & -- & -- & -- & -- & -- & -- \\
\cline{2-9}
& 6 & \checkmark & 0.834 & 27043000--27048000 & 27045500 & 1370 & 0 & \checkmark \\
\midrule

\multirow{6}{*}{%
  \parbox{3.9cm}{\raggedright SVM}
}
& 1 & \checkmark & 0.000 & 6813000--6818000 & 6815500 & 103095 & 100595 & \xmark \\
\cline{2-9}
& 2 & \checkmark & 1.250 & 6817500--6822500 & 6820000 & 1749 & 0 & \checkmark \\
\cline{2-9}
& 3 & \checkmark & 0.000 & 19318000--19323000 & 19320500 & 3560 & 1060 & \xmark \\
\cline{2-9}
& 4 & \checkmark & 0.000 & 22400500--22405500 & 22403000 & 2910 & 410 & \xmark \\
\cline{2-9}
& 5 & \xmark & -- & -- & -- & -- & -- & -- \\
\cline{2-9}
& 6 & \xmark & -- & -- & -- & -- & -- & -- \\
\bottomrule
\end{tabular}

\caption{Predicted coalescence intervals obtained with the classical
machine-learning models and their comparison with the true coalescence
times. The absolute midpoint error is the absolute difference between
the midpoint of the predicted interval and the true coalescence time.
The outside error is zero when the true coalescence time lies within
the predicted interval.}
\label{tab:classical_results_sangria}
\end{table*}

A comparison of the predicted merger lead times shows that the baseline VQC consistently identifies events approximately 0.3-1~h before coalescence. In contrast, several models frequently produce $\tau=0$, indicating that the highest confidence prediction occurs only immediately prior to the merger. Such behaviour limits the usefulness of these models for early warning applications despite their acceptable detection rates.

The model complexity comparison in Table \ref{tab:model_complexity_split} highlights that increased complexity does not necessarily improve $t_c$ estimation. The baseline VQC achieves the best localisation performance with only 162 trainable parameters and a circuit depth of 43. In contrast, the post VQC methods introduce substantially more classical processing overhead and, in some cases, significantly deeper quantum circuits without improving prediction accuracy. The most extreme example is the combined observable and ansatz expansion heuristic post VQC, which requires a circuit depth of approximately 2000 yet fails to detect any merger events. Similarly, among the classical models, the SVM possesses the largest number of trainable parameters but exhibits the poorest localisation behaviour, whereas the LSTM attains competitive performance with a substantially smaller model size. These observations suggest that the quality of the learned representation is more important than raw model complexity for merger-time prediction. Overall, the results indicate that the baseline VQC offers the most favourable trade-off between computational complexity, detection capability, and coalescence time localisation accuracy. Its ability to consistently produce intervals containing the true coalescence time while maintaining relatively modest circuit depth and parameter count makes it the most effective model among those considered in this study.

\begin{table*}[htbp]
\centering
\footnotesize
\setlength{\tabcolsep}{2pt}
\renewcommand{\arraystretch}{1.25}

\begin{tabularx}{\textwidth}{@{}
  >{\raggedright\arraybackslash}m{2.8cm}
  >{\centering\arraybackslash}m{1.05cm} 
  >{\centering\arraybackslash}m{1.05cm} 
  >{\centering\arraybackslash}m{0.85cm}@{}
  >{\centering\arraybackslash}m{0.85cm} 
  >{\raggedright\arraybackslash}X
  >{\raggedright\arraybackslash}X
@{}}
\toprule
\textbf{Model} &
\multicolumn{2}{c}{\textbf{\# params}} &
\multicolumn{2}{c}{\textbf{\shortstack{Quantum circuit\\depth}}} &
\multicolumn{2}{c}{\textbf{Time complexity}} \\
\cmidrule(lr){2-3}
\cmidrule(lr){4-5}
\cmidrule(lr){6-7}
& \textbf{GW detection} & \textbf{Merger detection}
& \textbf{GW detection} & \textbf{Merger detection}
& \textbf{Quantum circuit time} & \textbf{Classical processing time} \\
\midrule
VQC
& 162 trainable & 162 trainable & 43 & 43
& $O(N n 2^n L)$ & -- \\
\addlinespace[2pt]
Post VQC, observable construction heuristic
& 471 (classical) & 57 (classical) & 2 & 57
& $\begin{aligned}
O\Big( N 2^n &\sum_{l=1}^{n-1} l \Big( n \\
&+ \sum_{j=1}^{l} 3^j \tbinom{n}{j} \Big) \Big)
\end{aligned}$
& $\begin{aligned}
O\Big( N \Big( &\sum_{j=1}^{n-1} 3^j \tbinom{n}{j}\, h_1 \\
&+ \sum_{i=2}^{l} h_{i-1} h_i \Big) \Big)
\end{aligned}$ \\
\addlinespace[2pt]
Post VQC, ansatz expansion heuristic
& 4 fixed + 33 classical
& 4 fixed + 17 classical
& 8 & 8
& $O\Big( N n 2^n 2^o \tbinom{2n}{o} \Big)$
& $\begin{aligned}
O\Big( N \Big( &2^o \tbinom{2n}{o} h_1 \\
&+ \sum_i h_{i-1} h_i \Big) \Big)
\end{aligned}$ \\
\addlinespace[2pt]
Post VQC, observable construction heuristic + ansatz expansion heuristic
& 20 fixed + 49 classical
& 193 (classical) & 2000 & --
& $O\Big( N n 2^n 2^o \tbinom{2n}{o} \Big)$
& $\begin{aligned}
O\Big( N \Big( &\Big( \sum_{j=1}^{l} 3^j \tbinom{n}{j} 2^o \tbinom{2n}{o} \Big) h_1 \\
&+ \sum_i h_{i-1} h_i \Big) \Big)
\end{aligned}$ \\
\addlinespace[2pt]
Data Reuploading Classifier
& 3 trainable & 3 trainable & 6 & 6
& $O(N n 2^n L)$ & $O(N n 2^n L)$ \\
\addlinespace[2pt]
QLSTM
& 105 trainable & 36 trainable & 26 & 9
& $O(N L h n 2^n)$
& $\begin{aligned}
O\Big( NL \big( &\text{input dim}\cdot h \\
&+ h^2 + h n 2^n \big) \Big)
\end{aligned}$ \\
\addlinespace[2pt]
Quantum Kernel SVM
& -- & -- & 7 & --
& $O(N^2 n 2^n)$ & $O(N^3)$ \\
\midrule
SVM
& 2229 & 3296 & -- & -- & -- & $O(N^3)$ \\
\addlinespace[2pt]
MLP
& 262 & 69 & -- & -- & -- & $O\Big( N \sum_i h_{i-1} h_i \Big)$ \\
\addlinespace[2pt]
RNN
& 212 & 56 & -- & -- & -- & $O\big( N L (\text{input dim}\cdot h + h^2) \big)$ \\
\addlinespace[2pt]
LSTM NN
& 266 & 206 & -- & -- & -- & $O\big( N L (\text{input dim}\cdot h + h^2) \big)$ \\
\bottomrule
\end{tabularx}
\caption{Comparison of model complexity for GW detection and merger detection tasks.}
\label{tab:model_complexity_split}
\end{table*}

\section{Conclusion}

In this work, we benchmarked a diverse set of quantum, hybrid quantum--classical, and classical NN models for two tasks relevant to the analysis of LISA data: the detection of MBHB GW signals and the prediction of their coalescence times. All models were trained on the Sangria training dataset and subsequently evaluated on the Sangria blind dataset, providing a common framework for assessing their ability to generalise to previously unseen signals embedded in realistic LISA noise.

For GW signal detection, several quantum models achieved performance comparable to or exceeding that of the classical baselines. The post VQC architecture combining the observables and ansat expansion heuristics correctly identified all signal samples in the blind dataset, while the standard VQC and the data reuploading classifier achieved recalls of approximately $0.996$, outperforming all classical NN models and the quantum kernel SVM. At the same time, the results revealed substantial variation among the quantum architectures, demonstrating that the use of quantum circuits alone does not guarantee superior performance. Some quantum models, particularly the QLSTM and the quantum kernel SVM, were outperformed by the best classical baseline, namely the classical SVM.

For merger detection and coalescence time estimation, a different picture emerged. Although the post VQC approach combining both heuristics achieved the highest GW detection recall, it failed to successfully identify any merger events. In contrast, the standard VQC provided the most reliable coalescence localisation. It successfully detected five out of the six merger events present in the blind dataset and, for every detected event, produced coalescence intervals containing the true $t_c$. Moreover, it achieved the lowest midpoint errors among all investigated models while maintaining a comparatively modest circuit depth and number of trainable parameters. These results demonstrate that the baseline VQC provides the most favourable trade-off between detection performance, localisation accuracy, and model complexity.

The benchmark also highlights that increasing architectural complexity does not necessarily improve performance. Several post VQC models required substantially more classical processing and deeper quantum circuits while failing to outperform the simpler VQC architecture. A similar trend was observed among the classical models, where larger parameter counts were not systematically associated with better localisation accuracy. These results suggest that the quality of the learned representation is more important than model size alone for the analysis of GW signals.

Overall, the results provide evidence for a practical usefulness of QML models for LISA data analysis. In particular, the standard VQC and selected hybrid architectures achieved detection and localisation performances that were competitive with, and in some cases superior to, those of established classical methods. However, the present work does not constitute evidence of a provable quantum advantage in the complexity theoretic sense. All experiments were performed on small scale quantum simulations and involve feature engineered inputs that are also accessible to classical models. The observed improvements should therefore be interpreted as a quantum model advantage for specific tasks rather than a demonstration of computational superiority, constituting an application use case for QML in the domain of space data analysis.

Nevertheless, the ability of relatively compact quantum models to achieve competitive performance on realistic LISA data suggests that QML may become a valuable component of future GW analysis pipelines. As quantum hardware continues to mature, larger circuits, higher dimensional feature embeddings, and end-to-end quantum processing strategies may enable more expressive models and allow a more stringent assessment of possible quantum advantages in astrophysical signal processing. Future work will focus on testing these architectures on additional LISA Data Challenge datasets, incorporating parameter-estimation tasks, and evaluating their performance on noisy quantum hardware.

\section{Acknowledgements}
This work is supported by the Romanian Ministry of Research, Innovation and Digitalization under the Romanian National Core Program LAPLAS VII - contract no. 30N/2023 and by ESA PRODEX project RoLISASpace.

\bibliographystyle{plain}
\bibliography{bib}

\end{document}